\documentclass[twocolumn]{aastex63}

\usepackage{amssymb, amsfonts, amsmath,framed}
\usepackage{comment}
\usepackage{enumitem}

\usepackage{bm}
\expandafter\ifx\csname package@font\endcsname\relax\else
 \expandafter\expandafter
 \expandafter\usepackage
 \expandafter\expandafter
 \expandafter{\csname package@font\endcsname}
\fi
\expandafter\ifx\csname package@font\endcsname\relax\else
 \expandafter\expandafter
 \expandafter\usepackage
 \expandafter\expandafter
 \expandafter{\csname package@font\endcsname}
\fi
\def\bi{\begin{itemize}}
\def\ei{\end{itemize}}
\def\bq{\begin{equation}}
\def\eq{\end{equation}}
\def\bqy{\begin{eqnarray}}
\def\eqy{\end{eqnarray}}

\begin{document}
\title{Pitch-Angle Anisotropy Imprinted by Relativistic Magnetic Reconnection}

\correspondingauthor{}
\email{luca.comisso@columbia.edu}

\author{Luca Comisso}
\affiliation{Department of Astronomy and Columbia Astrophysics Laboratory, Columbia University, New York, New York 10027, USA}

\author{Brian Jiang}
\affiliation{Columbia College, Columbia University, New York, New York 10027, USA}

\begin{abstract}
Radiation emitted by nonthermal particles accelerated during relativistic magnetic reconnection is critical for understanding the nonthermal emission in a variety of astrophysical systems, including blazar jets, black hole coronae, pulsars, and magnetars. 
By means of fully kinetic Particle-in-Cell (PIC) simulations, we demonstrate that reconnection-driven particle acceleration imprints an energy-dependent pitch-angle anisotropy and gives rise to broken power laws in both the particle energy spectrum and the pitch-angle anisotropy. 
The particle distributions depend on the relative strength of the non-reconnecting (guide field) versus the reconnecting component of the magnetic field ($B_g/B_0$) and the lepton magnetization ($\sigma_0$). 
Below the break Lorentz factor $\gamma_0$ (injection), the particle energy spectrum is ultra-hard ($p_< < 1$), while above $\gamma_0$, the spectral index $p_>$ is highly sensitive to $B_g/B_0$. 
Particles' velocities align with the magnetic field, reaching minimum pitch angles $\alpha$ at a Lorentz factor $\gamma_{\min \alpha}$ controlled by $B_g/B_0$ and $\sigma_0$. The energy-dependent pitch-angle anisotropy, evaluated through the mean of $\sin^2 \alpha$ of particles at a given energy, exhibits power-law ranges with negative ($m_<$) and positive ($m_>$) slopes below and above $\gamma_{\min \alpha}$, becoming steeper as $B_g/B_0$ increases. 
The generation of anisotropic pitch angle distributions has important astrophysical implications. We address their effects on regulating synchrotron luminosity, spectral energy distribution, polarization, particle cooling, the synchrotron burnoff limit, emission beaming, and temperature anisotropy.

\vspace{0.9cm}

\end{abstract}

\section{Introduction} \label{sec:intro} 
Magnetic reconnection plays a pivotal role in astrophysical plasmas, enabling the topological reconfiguration of magnetic field lines and the rapid conversion of magnetic energy into kinetic energy \citep{Biskamp_MR2000,Kulsrud2005,Ji2022Nat}.
It is widely recognized as the key mechanism driving a variety of energetic phenomena, including solar flares \citep[e.g.][]{ShiMag11,GuoL20}, $\gamma$-ray emission from pulsar magnetospheres \citep[e.g.][]{Cerutti16,Hakobyan23}, flares in the Crab Nebula \citep[e.g.][]{Cerutti13,Lyut18}, and $\gamma$-ray flares from active galactic nuclei \citep[e.g.][]{Gia09,Sob23}. Indeed, it is believed that particle acceleration collisionless via magnetic reconnection powers the observed bright nonthermal radiation associated with these astrophysical phenomena.

In magnetically dominated astrophysical environments, the magnetic energy density available for reconnection can exceed not only the plasma pressure but also the rest mass energy density of the plasma. Under such conditions, the Alfv{\'e}n speed approaches the speed of light, promoting magnetic reconnection in the relativistic regime. While analytical studies have provided comprehensive insights into the dynamics of relativistic reconnection  \citep{LU03,Lyu05,CA14}, capturing the particle acceleration process necessitates first-principles numerical simulations. An extensive exploration of particle acceleration in relativistic reconnection has been undertaken through a multitude of Particle-in-Cell (PIC) studies \citep[e.g.][]{Zenitani01,Jaroschek04,Lyub08,Bessho12,Cerutti12,Kagan13,Sironi14,Guo14,Melzani14,Werner16}. It is now well-established that collisionless relativistic reconnection gives rise to nonthermal particle distributions when the reconnecting magnetic field is much larger than the non-reconnecting component of the magnetic field, known as the `guide field'.

The presence of even a moderate guide field can significantly alter the reconnection dynamics and the particle energization process. Recent simulations investigating collisionless magnetic reconnection within a turbulence cascade \citep{Comisso19,Comisso20ApJL,Comisso21,Comisso22} have uncovered that the guide field not only shapes the distribution of particle energies but also controls the distribution of pitch angles—namely, the angles between particle velocities and the local magnetic field. These simulations have demonstrated that reconnection within turbulence can induce a pronounced energy-dependent pitch-angle anisotropy, the extent of which depends on the strength of the guide field. 

Accurately predicting the pitch angle distribution of the emitting particles is of paramount importance for interpreting synchrotron radiation from high-energy sources such as pulsar wind nebulae \citep[e.g.][]{Hester08,Reynolds17,Lyutikov19}, magnetospheres of supermassive black holes \citep[e.g.][]{Johnson15,EHT_M87_19,EHT_SgrA_22}, jets from active galactic nuclei \citep[e.g.][]{Tavecchio18,deJaeger23,DiGesu23} or $\gamma$-ray bursts \citep[e.g.][]{Burgess14,Oganesyan17,Ravasio19,Burgess20}. However, the role of isolated reconnection layers in generating pitch angle anisotropy has remained unexplored.

In this paper, we investigate the simultaneous generation of energetic particles and pitch-angle anisotropy within collisionless reconnection layers through rigorous first-principles PIC simulations. We demonstrate that reconnection-driven particle acceleration results in energy-dependent pitch-angle anisotropy and broken power laws in both the particle energy spectrum and the pitch-angle anisotropy. Their properties are determined by the relative strength of the guide field compared to the reconnecting magnetic field. Additionally, spectral features, including breaks and power-law slopes, depend on both guide field strength and magnetization. Remarkably, the low-energy (injection) spectrum exhibits very weak sensitivity to guide field strength and magnetization, whereas spectral properties above the injection range display significant dependencies on these parameters.

This paper is organized as follows. 
In Section \ref{sec:method}, we describe our numerical approach and simulation setup. 
In Section \ref{sec:results}, we present the results of our fully kinetic simulations. This includes scalings of the reconnection rate, the self-consistent particle energy distributions, and the simultaneous energy-dependent pitch-angle anisotropy. We provide relationships for characterizing breaks in the power laws of the particle energy distributions, as well as breaks in the power laws of the energy-dependent pitch-angle anisotropy.  
In Section \ref{sec:implications}, we discuss several astrophysical implications of the energy-dependent pitch-angle anisotropy. Specifically, we examine the role of the pitch-angle anisotropy in regulating the synchrotron luminosity, the spectral energy distribution, polarization of synchrotron radiation, particle cooling, the synchrotron burnoff limit, emission beaming, and temperature anisotropy.
Finally, in Section \ref{sec:conclusions} we summarize our findings.

\section{Numerical Method and Setup} \label{sec:method} 
We solve the relativistic Vlasov-Maxwell system of equations employing the PIC method \citep{birdsall_langdon_85} with the publicly available code TRISTAN-MP \citep{buneman_93, spitkovsky_05}. 
Our simulations are carried out in a two-dimensional domain where we track all three components of the electromagnetic field and particle momenta. 
The computational domain is periodic in the $x$-direction and continually expands in the $y$-direction, where two moving injectors, receding from $y = 0$ at the speed of light, continuously introduce fresh magnetized plasma into the simulation domain (see \citet{Sironi14} for additional details). 

The plasma consists of electrons and positrons, with an ambient particle density of $n_0$. 
We initialize the particles based on a Maxwell-J\"{u}ttner distribution characterized by a thermal spread $\theta_0 = k_B T_0/m_e c^2$, where $T_0$ denotes the temperature, $k_B$ is the Boltzmann constant, $m_e$ is the electron mass, and $c$ is the speed of light. 
For the initial current sheet, we adopt a Harris sheet equilibrium with magnetic field given by $\bm{B} = B_0 \tanh{(y/\lambda)}\bm{\hat{x}}+B_g\bm{\hat{z}}$, where $B_0$ is the strength of the reconnecting magnetic field, $B_g$ is the strength of the guide field, and $\lambda$ is the half-thickness of the current sheet. The particle density profile across the current sheet follows $n = n_0 [1 + 3 \cosh^{-2}(y/\lambda)]$. Particles within the current sheet have a drift velocity in the $z$-direction, ensuring the current density satisfies Amp\'ere's law.

The initial current sheet is unstable to tearing modes \citep{FKR63}, as determined by the tearing stability index $\Delta' = (2/\lambda) [(k \lambda)^{-1} - k\lambda]$, where $k$ denotes the perturbation wavenumber. Tearing modes grow when $\Delta' > 0$, implying that the Harris sheet is unstable to modes with $k\lambda < 1$. 
We set $\lambda = 5 d_e$, where $d_e=c/\omega_p$ represents the plasma skin depth, $\omega_p$ is the plasma frequency defined as $\omega_p=(4 \pi n_0 e^2/m_e)^{1/2}$, and $e$ denotes the electron charge. To initiate reconnection, we perturb the Harris sheet equilibrium with a small-amplitude, long-wavelength perturbation. 
For the computational domain in the $x$-direction, we adopt a half-length of $L_x = 1200 d_e$. 
We employ a grid cell size of $\Delta x = 0.25 d_e$ and initialize 16 particles per cell for the ambient plasma. We set the numerical speed of light to 0.45 cells per time step, ensuring that the CFL condition is satisfied.

In our simulations, we adopt $\theta_0 = k_B T_0/ m_e c^2 = 0.03$. Our results are not sensitive to the initial dimensionless temperature $\theta_0$ except for an overall energy rescaling \citep{Comisso19}. 
The strength of the reconnecting magnetic field is parametrized by the lepton magnetization $\sigma_0 = B_0^2/4\pi n_0 m_e c^2$. The magnetization associated with the combined magnetic field is $\sigma = (\omega_L/\omega_p)^2= \sigma_0 + \sigma_g$, where $\omega_L=eB/m_e c$ represents the nonrelativistic Larmor frequency and $\sigma_g = B_g^2/4\pi n_0 m_e c^2$ corresponds to the magnetization associated with the guide field. 
To comprehensively explore the parameter space defined by the two fundamental parameters of the problem, $B_g$ and $\sigma_0$, we conduct 35 simulations, spanning 7 different values of the guide field, $B_g = (1/8, 1/4, 1/2, 1, 2, 4, 8) B_0$, and 5 different values of magnetization, $\sigma_0 = (4, 8, 16, 32, 64)$. 

In all simulations, the Alfv{\'e}n speed associated with the reconnecting field approaches the speed of light, $v_{A0} = c \sqrt{\sigma_0/(1+\sigma_0)} \simeq c$. Each simulation runs for a minimum duration of $t = 5 (L_x/c) \max [1, B_g/B_0]$, ensuring we have an adequate time range to assess the steady-state properties of the system. We conduct our analysis on a large subset of particles that were randomly selected and tracked over time in each PIC simulation. The analysis excludes particles initially set up in the current sheet, as their properties depend on the initialization choices.

\section{Results} \label{sec:results} 

\begin{figure}
\begin{center}
    \includegraphics[width=8.65cm]{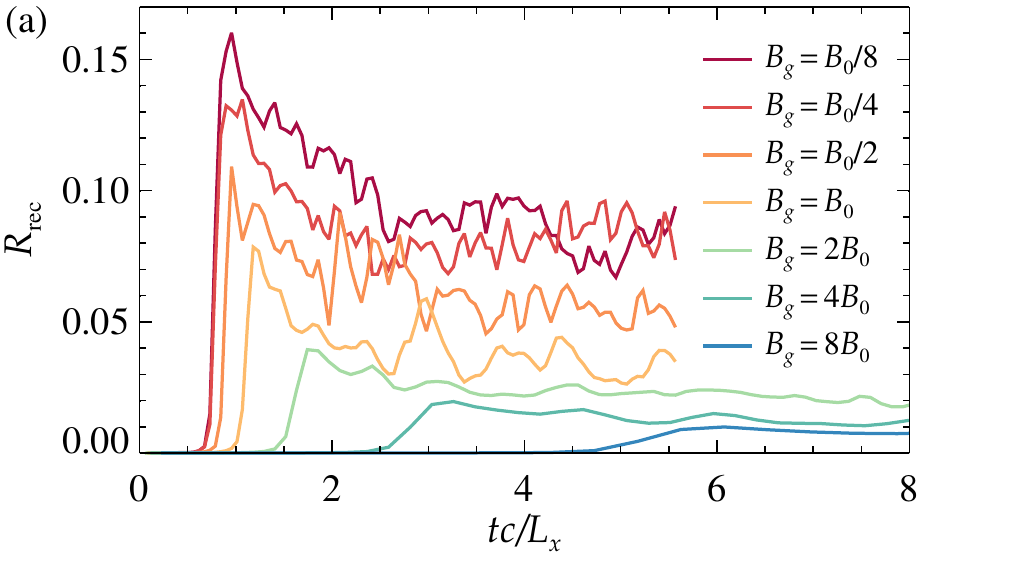}

\vspace{0.3cm}
    
    \includegraphics[width=8.65cm]{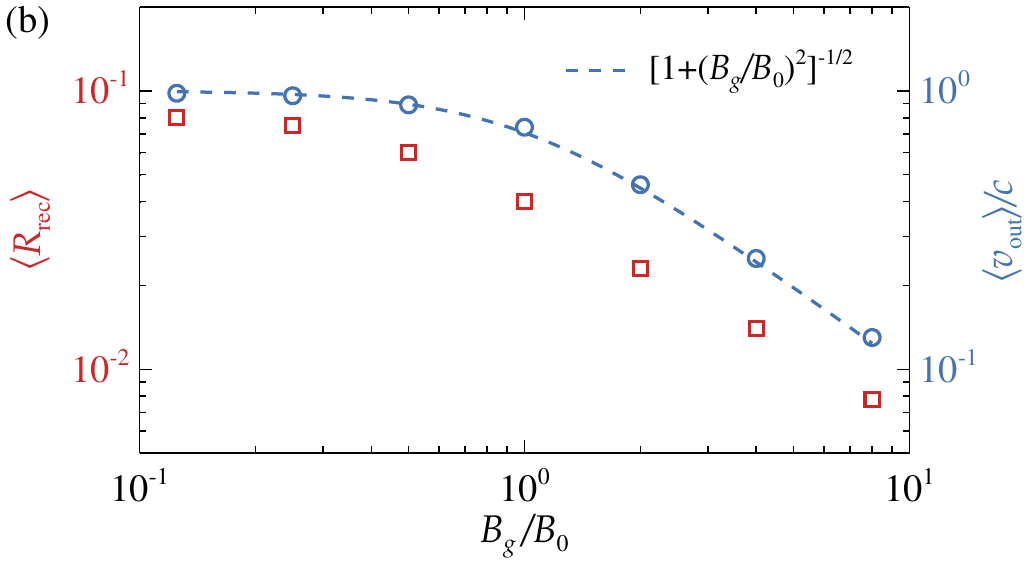}

\vspace{0.3cm}
    
    \includegraphics[width=8.65cm]{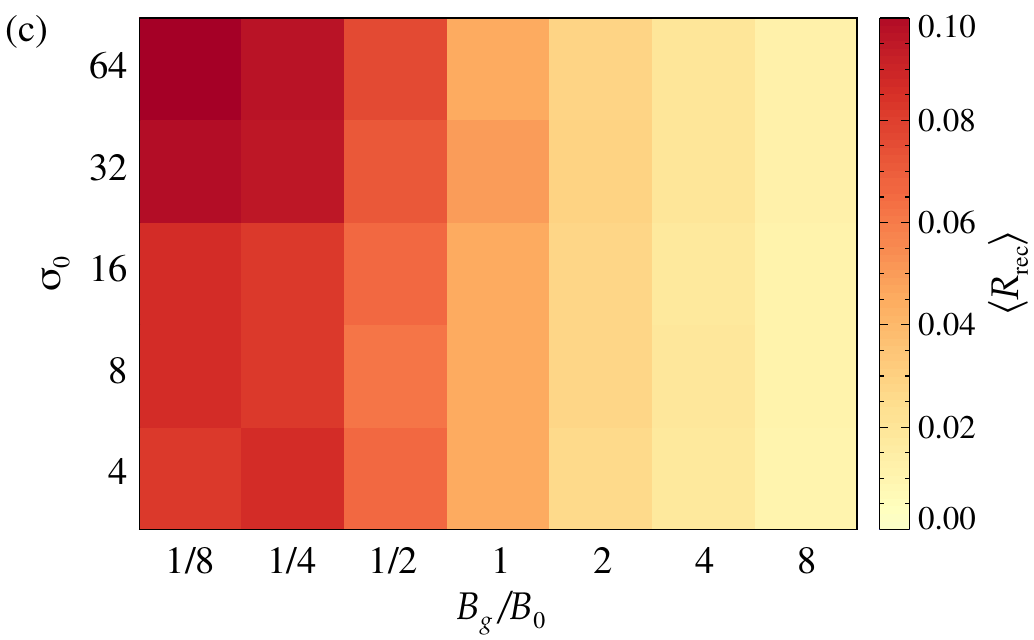}
\end{center}
\vspace{-0.2cm}
    \caption{(a) Time evolution of the reconnection rate $R_{\rm rec}$ for simulations with magnetization $\sigma_0 = 16$ spanning a broad range of guide field strengths $B_g$. (b) Time-averaged reconnection rate ($\langle R_{\rm rec} \rangle$) and maximum outflow velocity normalized to the speed of light ($\langle v_{\rm out} \rangle/c$) during the statistical steady-state regime ($t \gtrsim L_x/c$ after the peak of the reconnection rate) for the same simulations. (c) 2D histogram of the statistical steady-state reconnection rate $\langle R_{\rm rec} \rangle$ as a function of the dimensionless guide field strength $B_g/B_0$ and the magnetization $\sigma_0$.}
\label{fig1}
\end{figure}

In Figure \ref{fig1}(a), we illustrate how the reconnection rate evolves over time in simulations that vary in guide field magnitude. The normalized reconnection rate is calculated exactly as
\begin{equation}
   R_{\rm rec}(t) = \frac{1}{c B_0} \frac{\partial}{\partial t} \Big(\max(A_z) - \min(A_z)\Big) \, ,
\end{equation}
where $A_z$ is the $z$-component of the magnetic vector potential, evaluated at the resonant surface.
\begin{figure*}
    \centering
     \includegraphics[width=18.4cm]{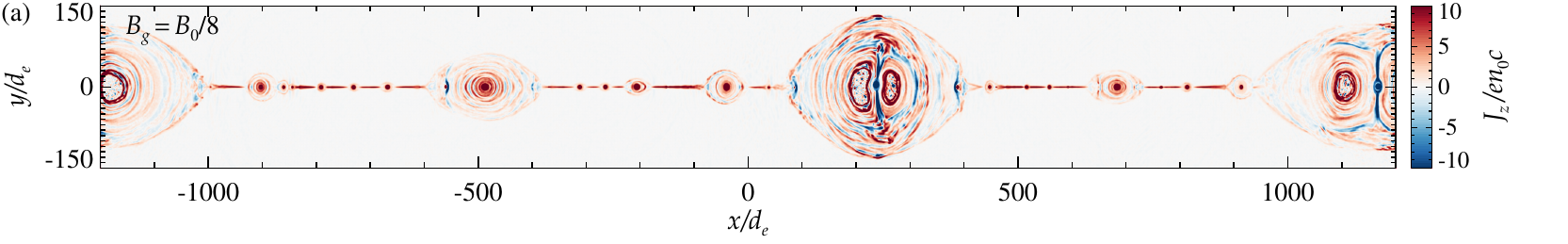}
     \includegraphics[width=18.4cm]{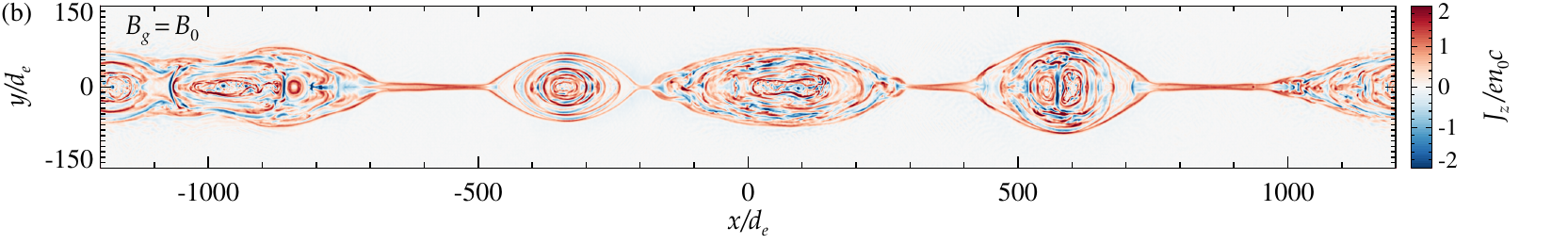}
     \includegraphics[width=18.4cm]{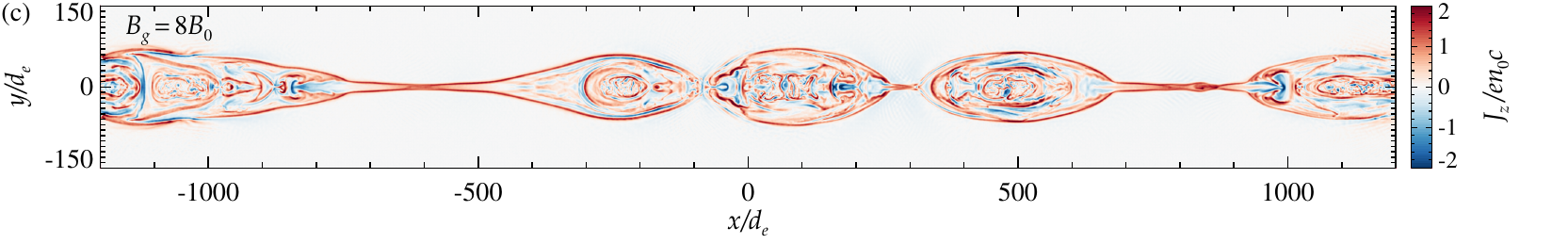}
     \includegraphics[width=18.4cm]{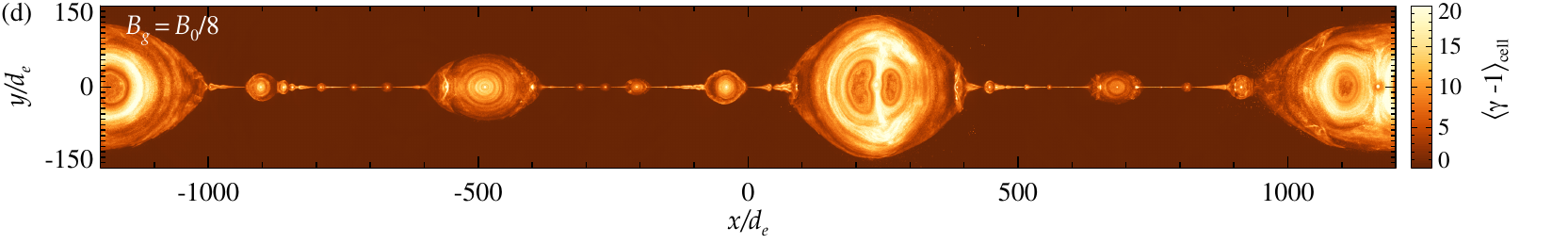}
     \includegraphics[width=18.4cm]{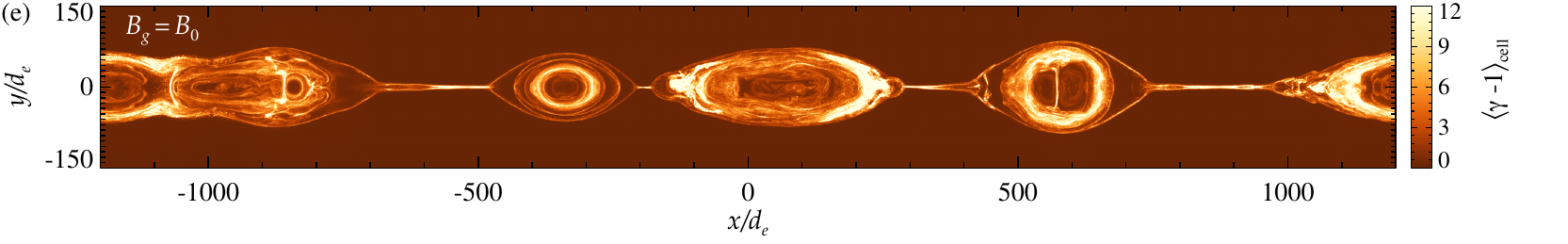}
     \includegraphics[width=18.4cm]{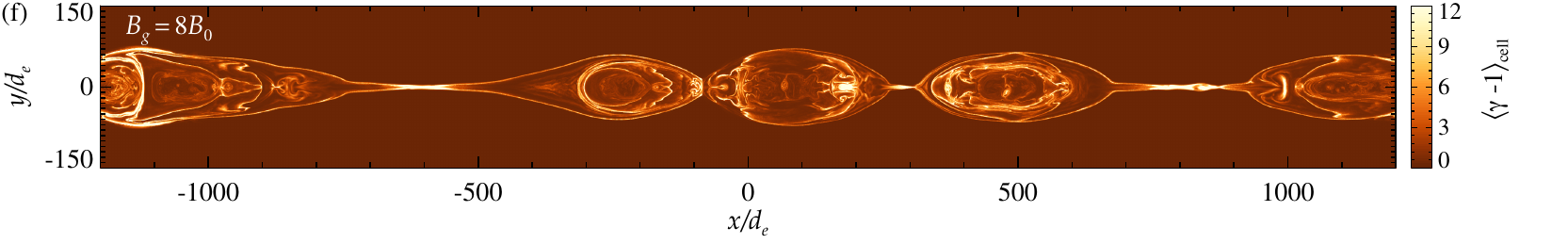}
    \caption{Snapshots of the electric current density and the mean kinetic energy per particle in the statistical steady-state regime for magnetic reconnection with varying guide field strength. The top-to-bottom panels correspond to guide field strengths of (a) $B_g=B_0/8$, (b) $B_g=B_0$, and (c) $B_g=8B_0$ for the out-of-plane current density $J_z$ (in units of $e n_0 c$), while panels (d), (e), and (f) illustrate the cell-averaged mean kinetic energy per particle $\langle \gamma - 1 \rangle_{\rm cell}$ (normalized by $m_e c^2$) for the same guide field strengths. The snapshots are taken at times around $t \sim 2 L_x (B_0^2+B_g^2)^{1/2}/B_0 c$, corresponding to similar stages of nonlinear evolution. To emphasize the small-scale structures in the reconnection layer, we limit the displayed domain region to $|y| \leq 150 d_e$.}
    \label{fig2}        
\end{figure*} 
After the initial linear growth phase preceding the onset of fast reconnection, the reconnection rate approaches a statistical steady state. Figure \ref{fig1}(a) shows that the reconnection rate decreases as the guide field strength increases. In particular, the reconnection rate, when time-averaged in the near steady state, is approximately
\begin{equation}
   \langle R_{\rm rec} \rangle \simeq 0.1 \, \frac{B_0}{({B_0^2 + B_g^2})^{1/2}} \, .
\end{equation}
The drop in the reconnection rate with increasing guide field strength results from the reduced outflow velocity in the reconnection layer. Indeed, in approximate steady state, the reconnection rate $R_{\rm rec} = E_{\rm rec}/B_0 \simeq v_{\rm in}/c \simeq 0.1 v_{\rm out}/c$ \citep{ComissoJPP16,CassakJPP17} scales with the outflow velocity. 
Given that the projection of the Alfv\'en speed into the outflow direction is $v_A \cos \zeta$, with $\zeta = \arctan (B_g/B_0)$, the outflow velocity in the statistical steady state is expected to decrease with the guide field strength as
\begin{equation} \label{eq:vout}
   \langle v_{\rm out} \rangle \simeq \frac{c B_0}{({B_0^2 + B_g^2})^{1/2}} \, .
\end{equation}
This is indeed confirmed by the scaling shown in Figure \ref{fig1}(b). 
Finally, in Figure \ref{fig1}(c), we present the statistical steady-state reconnection rate measured throughout our entire simulation campaign. In the $\sigma_0 \gg 1$ regime of interest here, the reconnection rate becomes largely independent of the magnetization parameter $\sigma_0$, while maintaining its strong dependence on the strength of the guide field.

In Figure \ref{fig2}, we show representative snapshots of the reconnection layer during the highly nonlinear phase of the reconnection process for simulations with $\sigma_0=16$ and varying guide field strength, namely $B_g = (1/8, 1, 8) B_0$. The reconnection layer undergoes fragmentation, resulting in the formation of secondary current sheets and plasmoids \citep{Comisso2016PoP,Comisso2017ApJ,Uzd2016}, where plasmoids are the distinctive island-like structures that form within the reconnecting current sheet. This phenomenon is depicted in panels (a)-(c), which display the electric current density in the out-of-plane direction. 
As the guide field strength increases, both the number of plasmoids and their hierarchical scale separation diminish. Furthermore, the inter-plasmoid current sheets transition to an open Petschek-like configuration \citep{Comisso2013}. This open, $X$-type configuration (as opposed to the $Y$-type Sweet-Parker-like geometry) effectively inhibits the formation of further plasmoids in the presence of a strong guide field (compare panel (c) with panel (a)).
In panels (d)-(f), we show the cell-averaged kinetic energy per electron. With increasing guide field strength, high-energy particle regions become more localized, predominantly aligning with $X$-points and plasmoid boundaries.

\begin{figure}
\begin{center}
    \includegraphics[width=8.65cm]{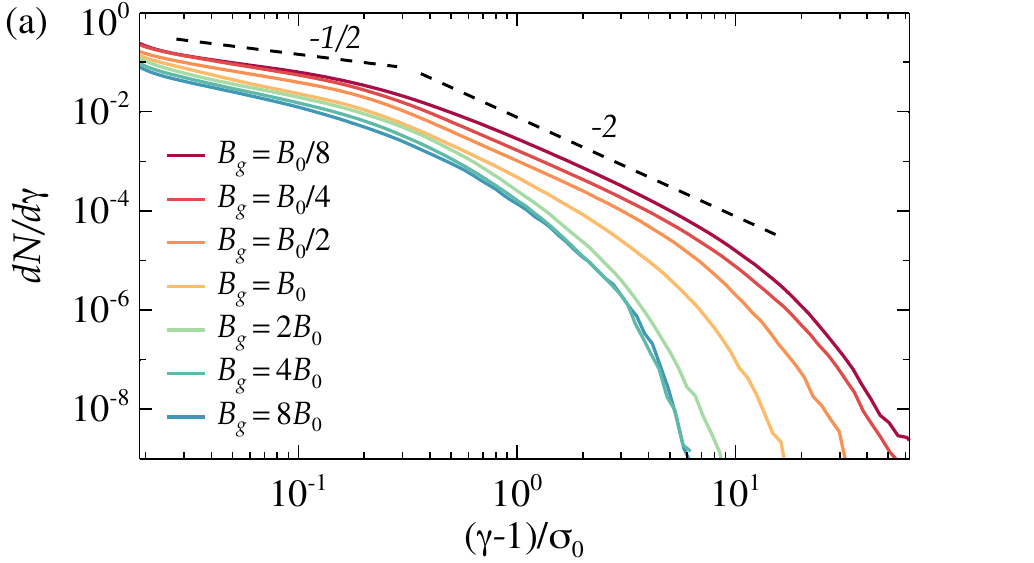}

\vspace{0.3cm}
    
    \includegraphics[width=8.65cm]{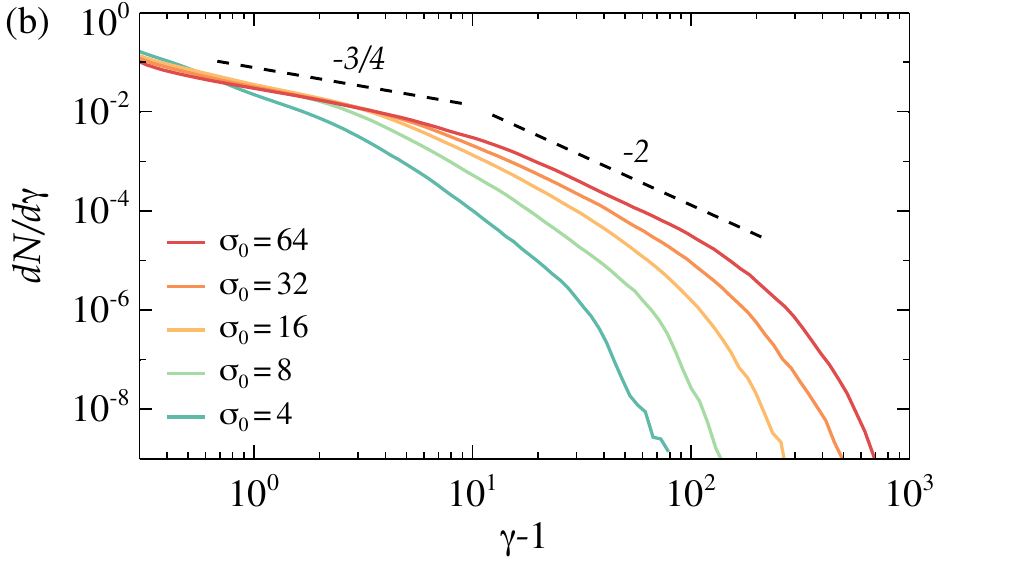}
\end{center}
\vspace{-0.2cm}
    \caption{Particle energy spectra $dN/d\gamma$ at late times, $t \simeq 3 L_x (B_0^2+B_g^2)^{1/2}/B_0 c$, when the spectra have fully developed. Panel (a) showcases simulations with fixed $\sigma_0 = 16$ and varying guide field strengths $B_g = (1/8, 1/4, 1/2, 1, 2, 4, 8) B_0$, while panel (b) features simulations with fixed $B_g/B_0 = 1$ and varying lepton magnetization $\sigma_0 = (4, 8, 16, 32, 64)$. Dashed lines indicating power-law slopes are provided for reference.}
\label{fig3}
\end{figure}

\begin{figure}
\begin{center}
    \includegraphics[width=8.65cm]{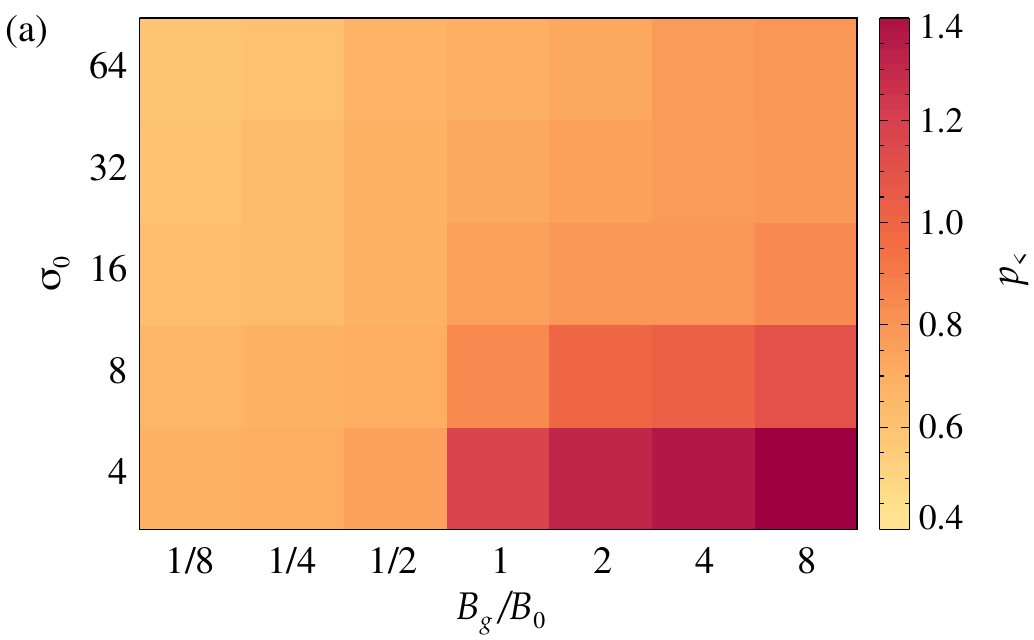}

\vspace{0.3cm}
    
    \includegraphics[width=8.65cm]{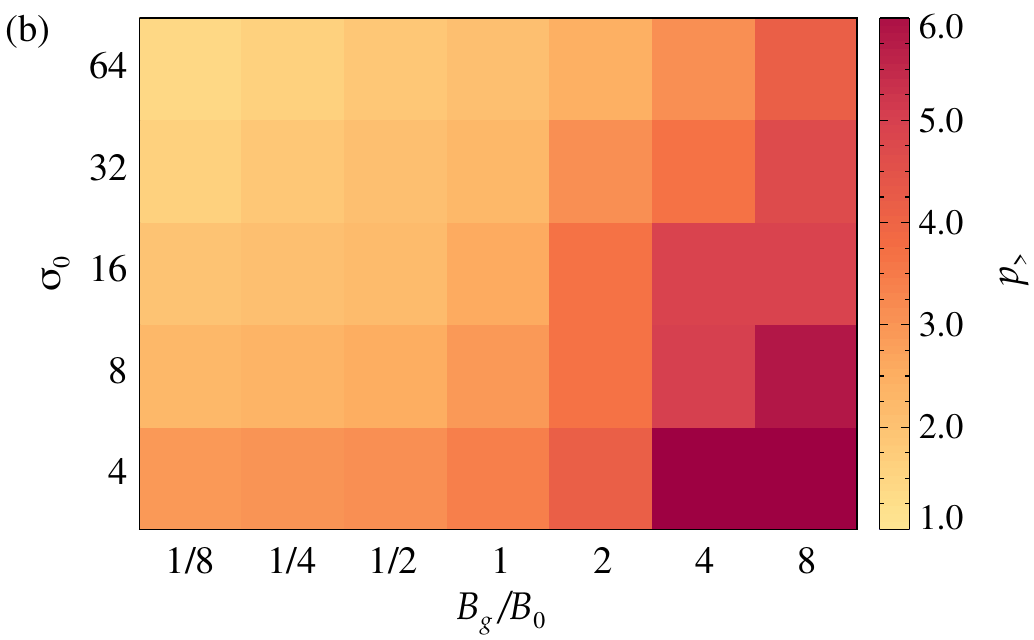}
\end{center}
\vspace{-0.2cm}
    \caption{2D histograms illustrating the power law indices $p_<$ (panel a) and $p_>$ (panel b), characterizing the nonthermal tail of the particle energy spectrum produced by magnetic reconnection, as a function of the dimensionless guide field strength $B_g/B_0$ and the magnetization $\sigma_0$.}
\label{fig4}
\end{figure}

Particles energized by reconnection display pronounced nonthermal tails, as illustrated in  Figure \ref{fig3}. 
In particular, Fig. \ref{fig3}(a) shows electron energy spectra integrated over the entire reconnection region (specifically for $|y| \leq 0.25 L_x$) for $\sigma_0 = 16$ with varying guide field strength, while Fig. \ref{fig3}(b) displays analogous energy spectra for $B_g = B_0$ with varying lepton magnetization.
These spectra are evaluated at a developed stage, about $t_{\rm dev} \simeq 3 L_x/\langle v_{\rm out} \rangle \simeq 3 L_x ({B_0^2 + B_g^2})^{1/2}/B_0 c$ from the onset of fast reconnection. 
A distinct characteristic of these particle energy spectra is their broken power-law structure. Specifically, for lepton magnetization $\sigma_0 \gg 1$ and relativistic particles, the particle energy spectra follow 
\begin{equation}
\label{eq:dNdgamma}
{N(\gamma)}{d\gamma} = 
\begin{cases}
K (\gamma/\gamma_0)^{-p_<} {d\gamma} \;  , &  \quad \gamma_{\rm th}<\gamma<\gamma_{0} \\ 
K (\gamma/\gamma_0)^{-p_>} {d\gamma} \;  , & \quad \gamma_{0}<\gamma<\gamma_{\rm cut} 
\end{cases}
\end{equation} 
Here, $\gamma_0$ is the break Lorentz factor that separates the two power law ranges, with $p_<$ and $p_>$ indicating the power-law indices for the lower and upper ranges, respectively. Additionally, $\gamma_{\rm th}$ and $\gamma_{\rm cut}$ denote the thermal Lorentz factor and high-energy cutoff Lorentz factor, while $K$ is a normalization constant. 

Our simulations indicate that $\gamma_0 = \kappa \sigma_0$, with $\kappa \simeq 0.2$, as depicted in Fig. \ref{fig3}(a). 
Notably, $\kappa$ exhibits minimal variation with respect to the guide field, with only a slight increase as $B_g$ increases. It remains within a two-fold range over the entire scan from $B_g = B_0/8$ to $B_g = 8 B_0$. 
On the other hand, an increase in $B_g/B_0$ significantly reduces the normalization constant $K$, as evidenced in Fig. \ref{fig3}(a). 
Indeed, the number efficiency, defined as the fraction of electrons in the reconnection region (the area between the reconnection separatrices) with $\gamma \geq \gamma_0$, decreases as $B_g$ increases—from $\sim 50\%$ when $B_g=B_0/8$ to $\sim 25\%$ when $B_g=B_0$, and further to $\sim 15\%$ when $B_g=8B_0$. 

Our simulations also indicate that $\gamma_{\rm cut} \simeq 4 \sigma_0$  when $B_g/B_0 > 1$. This threshold was previously proposed by \citet{Werner16} for reconnection in the absence of a guide field ($B_g=0$). In contrast, our simulations indicate that $\gamma_{\rm cut}$ grows as $B_g$ decreases. When $B_g/B_0 \ll 1$, $\gamma_{\rm cut}$ approaches the Hillas limit \citep{Hillas84}, which is determined by the particle gyroradius matching the width of the reconnection layer, namely $r_g(\gamma_{\rm cut}) \simeq \langle R_{\rm rec} \rangle 2 L_x$ \citep[see also][]{ZhangH21,LiX23}, which implies $\gamma_{\rm cut} \simeq 2 \langle R_{\rm rec} \rangle \sqrt{\sigma_0} L_x/d_e$.

Below the break Lorentz factor $\gamma_0$, particle energy spectra are extremely hard ($p_< < 1$). 
In simulations with $\sigma_0 = 16$, as displayed in Fig. \ref{fig3}(a), $p_< \simeq 1/2$ when $B_g/B_0 \ll 1$.  
In simulations with $B_g = B_0$, as illustrated in Fig. \ref{fig3}(b), $p_< \simeq 3/4$ when $\sigma_0 > 8$. The \emph{ultra-hard slope $p_< < 1$} arises because $\Delta \epsilon_e = \kappa \sigma_0 m_e c^2$ is the typical electron injection energy, leading to $\gamma dN/d\gamma$ peaking around $\gamma_0$. 
Above $\gamma_0$, the spectrum steepens, and the spectral index \emph{$p_>$ is guide field-sensitive} \citep[see also][]{LiX23}, in addition to depending on magnetization, as shown in Figs. \ref{fig3}(a) and \ref{fig3}(b). 
In comparison to the spectrum above $\gamma_0$ \citep[e.g.][]{Zenitani01,Jaroschek04,Lyub08,Sironi14,Guo15ApJ,Werner2017ApJL,Hakobyan21,LiX23,Zhang23ApJL}, the spectrum below $\gamma_0$ has not received significant attention so far. However, given that $\gamma_0 \simeq \kappa \sigma_0$, this spectral energy range can be substantial when $\sigma_0 \gg 1$. 
In the context of an electron-ion plasma, even as the overall magnetization (including both electrons and ions) approaches unity, $\sigma_0$ remains significantly large. This is because $2\sigma_0 = \sigma_{0,e} \simeq (m_i/m_e) \sigma_{0,i}$ where $m_i$ is the ion mass, while $\sigma_{0,e}$ and $\sigma_{0,i}$ correspond to the magnetization associated with electrons and ions, respectively.

Figure \ref{fig4} provides a comprehensive overview of the spectral indices derived from our simulation campaign.
The spectral index $p_<$, shown in Figure \ref{fig4}(a), exhibits a weak dependence on $B_g/B_0$ and becomes more sensitive to  $\sigma_0$ only when $\sigma_0$ falls outside the asymptotically large range and $B_g/B_0 \gtrsim 1$. When $\sigma_0 \gg 1$, $p_<$ consistently maintains its hardness, with $p_< < 1$, across all values of $B_g/B_0$ covered in our study.
In contrast, $p_>$ exhibits a stronger dependence on $B_g/B_0$ and a milder sensitivity to $\sigma_0$, as depicted in Figure \ref{fig4}(b). The ratio $B_g/B_0 \sim 1$ marks the transition from $p_> < 3$ to $p_> > 3$. When $B_g/B_0 \ll 1$, typical $p_>$ values hover around $p_> \simeq 2$. In contrast, when $B_g/B_0 \gg 1$, typical $p_>$ values tend to $p_> \sim 5$, with $p_> \sim 6$ for the lowest $\sigma_0$ values considered in our simulation campaign.

\begin{figure}
\begin{center}
    \includegraphics[width=8.65cm]{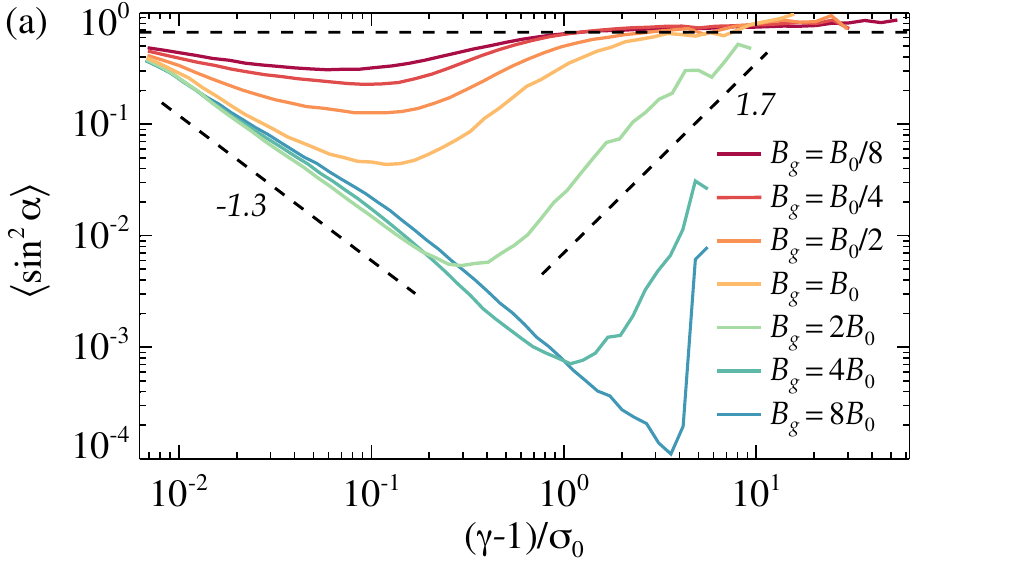}

\vspace{0.3cm}
    
    \includegraphics[width=8.65cm]{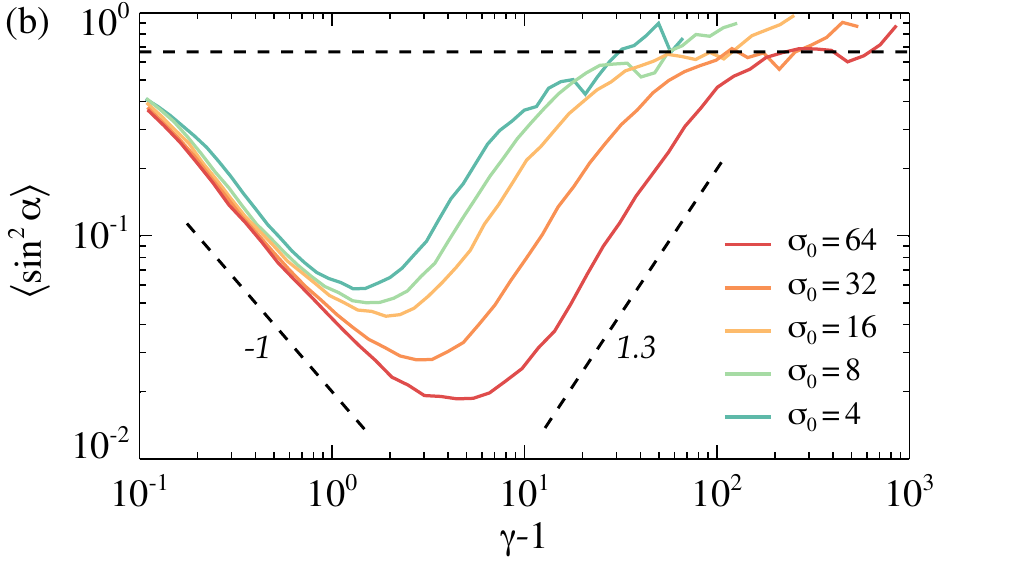}
\end{center}
\vspace{-0.2cm}
    \caption{Mean of the squared pitch angle sine, $\langle \sin^2 \alpha \rangle$, at late times, $t \simeq 3 L_x (B_0^2+B_g^2)^{1/2}/B_0 c$, for simulations with (a) fixed $\sigma_0 = 16$ and varying guide field strengths $B_g = (1/8, 1/4, 1/2, 1, 2, 4, 8) B_0$ and (b) fixed $B_g/B_0 = 1$ and varying lepton magnetization $\sigma_0 = (4, 8, 16, 32, 64)$. Horizontal dashed black lines indicate the expectation for isotropic particles, $\langle \sin^2 \alpha \rangle = 2/3$. In addition, dashed lines indicating power-law slopes are provided for reference.}
\label{fig5}
\end{figure}

\begin{figure}
\begin{center}
    \includegraphics[width=8.65cm]{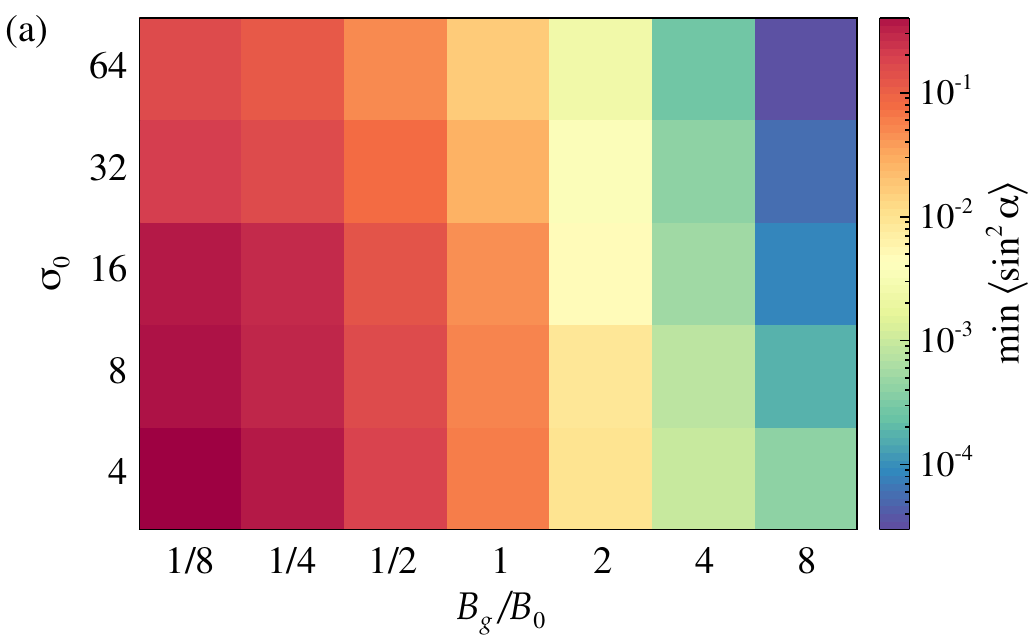}

\vspace{0.3cm}    
    
    \includegraphics[width=8.65cm]{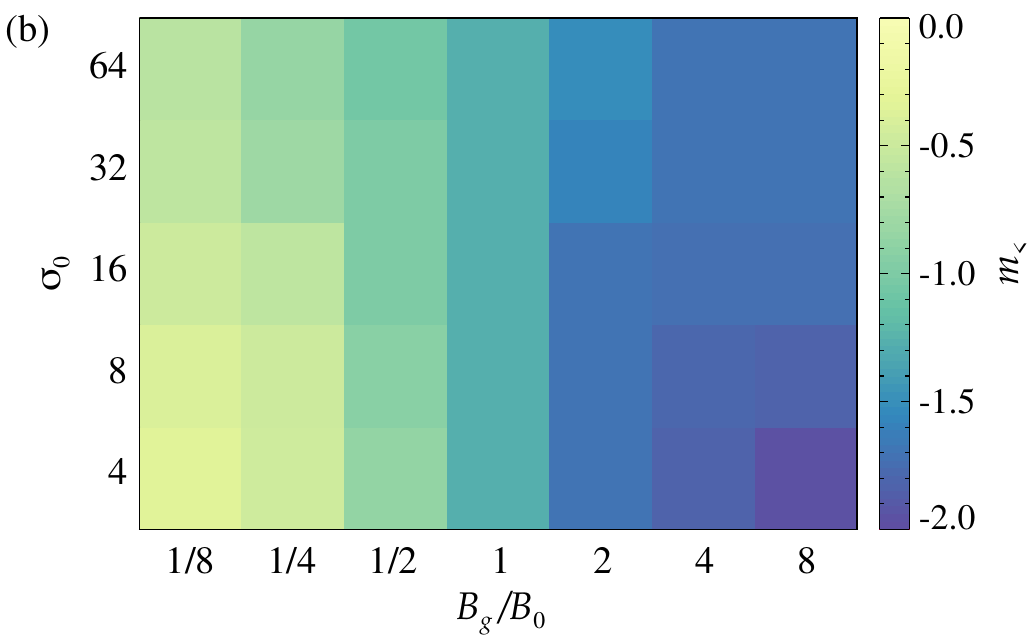}

\vspace{0.3cm}
    
    \includegraphics[width=8.65cm]{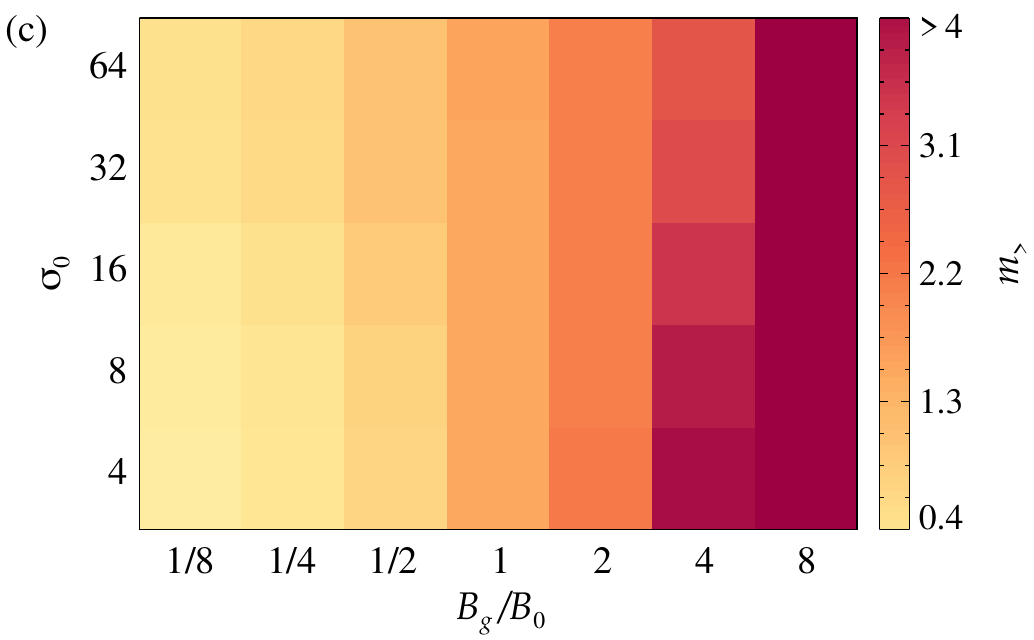}
\end{center}
\vspace{-0.2cm}
    \caption{2D histograms illustrating the peak anisotropy $\min \langle \sin^2 \alpha \rangle$ (panel a), along with the power law indices $m_<$ (panel b) and $m_>$ (panel c), describing the pitch angle distribution produced by magnetic reconnection, as a function of the dimensionless guide field strength $B_g/B_0$ and the magnetization $\sigma_0$.}
\label{fig6}
\end{figure}
Previous studies of magnetic reconnection within a turbulence cascade \citep{Comisso19,Comisso20ApJL,Comisso21,Comisso22} have shown that particle energization in the presence of a guide field gives rise to an energy-dependent pitch-angle anisotropy. 
To quantitatively assess this effect within an isolated reconnection layer, we examine the mean value of $\sin^2 \alpha$ as a function of $\gamma-1$, where $\alpha$ is the pitch angle, i.e. the angle between the particle momentum and the local magnetic field. Both $\alpha$ and $\gamma$ are measured in the local $\bm{E} \times \bm{B}$ frame. Here, we investigate the pitch-angle anisotropy across different $B_g/B_0$ and $\sigma_0$ values. 

Figure \ref{fig5} shows $\langle \sin^2 \alpha \rangle$ \emph{is energy-dependent}. Distinct power-law behaviors of $\langle \sin^2 \alpha \rangle$ are identifiable within specific energy ranges. 
For lepton magnetization $\sigma_0 \gg 1$ and relativistic particles, $\langle \sin^2 \alpha \rangle$ approximately follow
\begin{equation}
\label{eq:alphagamma}
\langle \sin^2 \alpha \rangle = 
\begin{cases}
\Lambda \left({\gamma}/{\gamma_{\min \alpha}}\right)^{m_<}  \;  , &  \quad \gamma_{\rm th}<\gamma<\gamma_{\min \alpha} \\ 
\Lambda \left({\gamma}/{\gamma_{\min \alpha}}\right)^{m_>}  \;  , &  \quad \gamma_{\min \alpha}<\gamma<\gamma_{\rm iso} \\ 
2/3 \;  , & \quad \gamma_{\rm iso}<\gamma<\gamma_{\rm cut} 
\end{cases}
\end{equation}
where $m_<$ and $m_>$ are the power-law indices characterizing the negative-slope and positive-slope energy dependence of the pitch-angle anisotropy, respectively, $\Lambda \sim \min \langle \sin^2 \alpha \rangle$, $\gamma_{\min \alpha}$ is the Lorentz factor at which the pitch-angle anisotropy is strongest, and $\gamma_{\rm iso}$ is the Lorentz factor at which particles return to a state close to pitch-angle isotropy.

We can identify three distinct regimes, contingent on $B_g/B_0$: 
\begin{enumerate}
\item $B_g/B_0 \ll 1$: the deviation of $\langle \sin^2 \alpha \rangle$ from the isotropic expectation of $2/3$ is generally modest (although it depends on $\sigma_0$, as discussed later), and $\min \langle \sin^2 \alpha \rangle$ is reached at $\gamma_{\min \alpha} \ll \sigma_0$, while pitch-angle isotropy is restored at $\gamma_{\rm iso} \sim \sigma_0 \ll \gamma_{\rm cut}$ (see Fig. \ref{fig5}(a)).
\item $B_g/B_0 \sim 1$: $\langle \sin^2 \alpha \rangle \ll 2/3$ and steep power-law slopes $m_<$ and $m_>$ manifest over a particle energy range characterized by $\gamma_{\min \alpha} \sim \gamma_0$ and $\gamma_{\rm iso} \sim 4 \sigma_0 < \gamma_{\rm cut}$. Accordingly, $\min \langle \sin^2 \alpha \rangle$ scales inversely with $\sigma_0$, and the extension of the rising range of $\langle \sin^2 \alpha \rangle$ is $\gamma_{\rm iso}/\gamma_{\min \alpha} \sim 4 \sigma_0/\gamma_0 \sim 20$ (see Figure \ref{fig5}(b)). 
\item $B_g/B_0 \gg 1$: $\langle \sin^2 \alpha \rangle$ displays the strongest deviations from isotropy, and $\gamma_{\min \alpha} \sim \gamma_{\rm cut} \sim 4 \sigma_0$. In this regime $\gamma_{\rm iso} \gg \gamma_{\rm cut}$, i.e. high-energy particles cannot undergo effective pitch angle isotropization. In this case, pitch angle anisotropy is characterized by an extended power law with index $m_<$ (see Figure \ref{fig5}(a)). 
\end{enumerate}

In Figure \ref{fig6}, we report the minimum of $\langle \sin^2 \alpha \rangle$ (panel a), along with the power-law slopes $m_<$ (panel b) and $m_>$ (panel c), obtained from the complete simulation campaign. The maximum anisotropy, for which $\min \langle \sin^2 \alpha \rangle$ serves as a proxy, increases with higher values of $B_g/B_0$ and $\sigma_0$. For the most extreme case in Figure \ref{fig6}(a), $B_g/B_0=8$ and $\sigma_0=64$, we obtain $\langle \sin^2 \alpha \rangle \sim 4\times 10^{-5}$, four orders of magnitude below the isotropic expectation. A demarcation from moderate to very strong depletion of $\min \langle \sin^2 \alpha \rangle$ occurs at $B_g/B_0 \sim 1$. Lower values of $\min \langle \sin^2 \alpha \rangle$ are clearly accompanied by steeper power-law ranges, as shown in Figures \ref{fig6}(b) and \ref{fig6}(c). The power-law slopes $m_<$ and $m_>$ exhibit a significantly more pronounced dependence on $B_g/B_0$ than on $\sigma_0$. Notably, in the regime $B_g/B_0 \sim 1$, $m_<$ and $m_>$ are almost independent of $\sigma_0$ within the parameter range explored in this study.

\begin{figure}
\begin{center}
    \includegraphics[width=8.65cm]{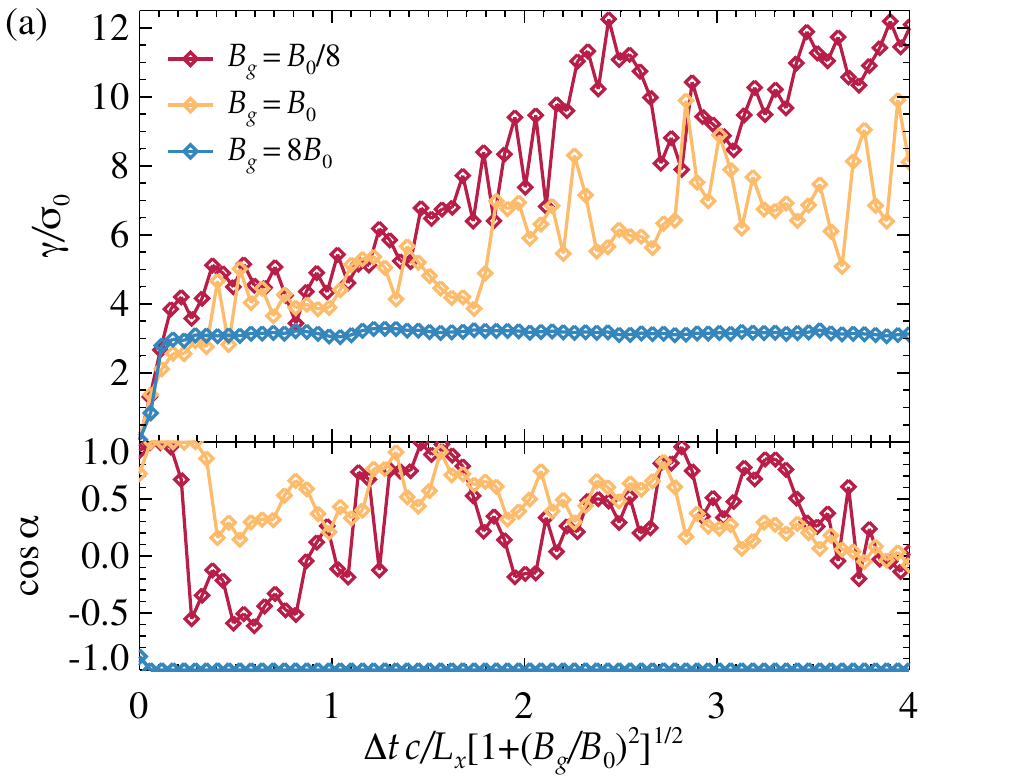}

\vspace{0.3cm}
    
    \includegraphics[width=8.65cm]{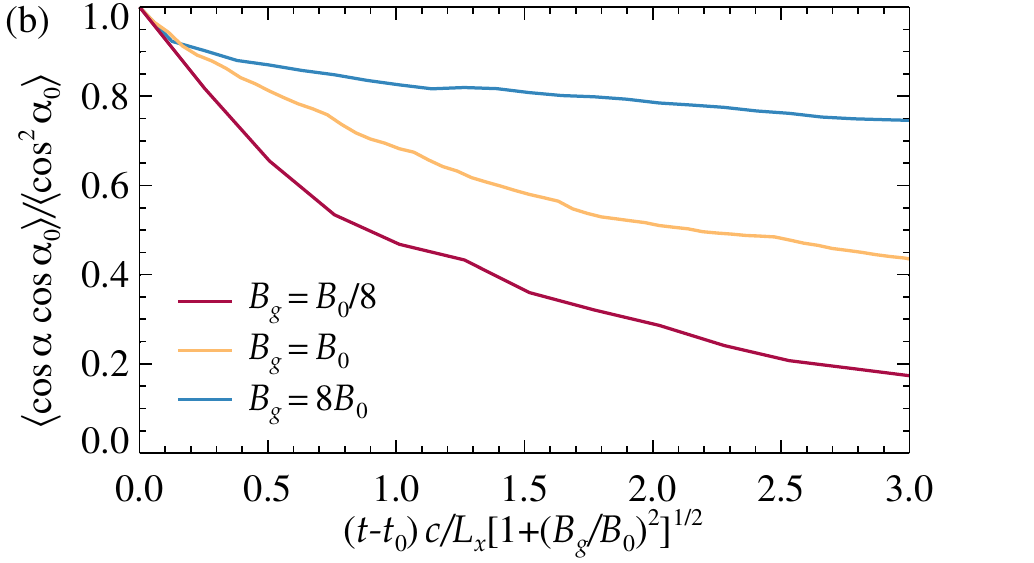}

\vspace{0.3cm}
    
    \includegraphics[width=8.65cm]{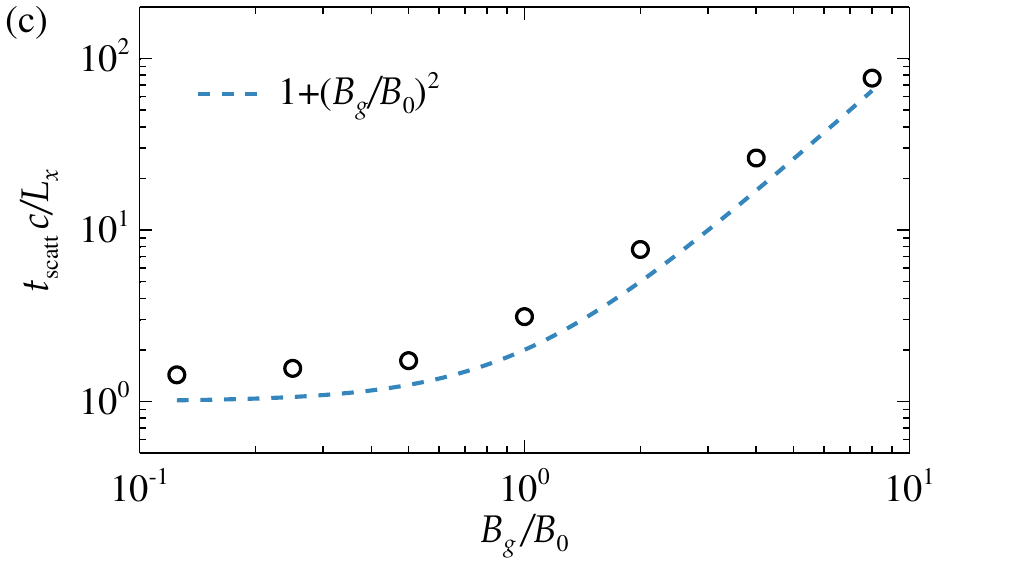}
\end{center}
\vspace{-0.2cm}
    \caption{(a) Examples of time evolutions of $\gamma/\sigma_0$ and $\cos{\alpha}$ from simulations with $\sigma_0=16$ and guide field strengths $B_g=B_0/8$ (red), $B_g=B_0$ (yellow), and $B_g=8 B_0$ (blue). The $x$-axis is scaled by $\Delta t (c/L_x) [1+(B_g/B_0)^2]^{-1/2}$, where $\Delta t$ corresponds to the time interval from the particle injection from the thermal pool ($\gamma \simeq \gamma_{\rm th}$) to higher Lorentz factors ($\gamma \gg \gamma_{\rm th}$). (b) Pitch-angle correlation function $\langle \cos{\alpha} \cos{\alpha_0} \rangle$ (normalized by $\langle \cos^2{\alpha_0} \rangle$), computed from the same simulations. $t_0$ indicates the initial time for the measurement. (c) Pitch-angle scattering timescale $t_{\rm scatt}$ (normalized by $L_x/c$), determined by fitting the pitch-angle correlation function with the exponential $e^{-t/t_{\rm scatt}}$, for simulations with $\sigma_0 = 16$ and various $B_g$ values. A dashed blue line indicating $1+(B_g/B_0)^2$ is provided for reference.}
\label{fig7}
\end{figure}

The emergence of pitch angle anisotropy can be understood in terms of a two-stage process, in line with previous studies of magnetic reconnection within a turbulence cascade \citep{Comisso18,Comisso19,Comisso21,Comisso22}. 
Different acceleration mechanisms, such as direct acceleration by the reconnection electric field \citep{Litvinenko96}, Fermi reflection \citep{Drake06}, or pickup acceleration \citep{Mobius85}, may operate in the initial stage of particle acceleration. However, as the guide field strength increases, direct acceleration by the reconnection electric field becomes progressively more important \citep{Dahlin16,French23}.
When the initial stage of particle acceleration occurs mostly in the magnetic field-aligned direction via the reconnection electric field, particles undergo an increase in their Lorentz factor as given by 
${d\gamma}/dt \simeq (e E_{\rm rec}/{m_e c}) \cos{\alpha} \simeq \langle R_{\rm rec}\rangle  B_0 (B_0^2+B_g^2)^{-1/2} \omega_L \cos{\alpha}$ (where we assumed $v \simeq c$). 
Simultaneously, the parallel component of the momentum equation of a particle in a constant parallel electric field yields ${d \cos{\alpha}}/dt \simeq ({e E_{\rm rec}}/{\gamma m_e c}) \sin^2{\alpha} \simeq \langle R_{\rm rec}\rangle B_0 (B_0^2+B_g^2)^{-1/2} ({\omega_L}/{\gamma}) \sin^2{\alpha}$, resulting in a reduction of the pitch angle for the accelerated particles. This initial acceleration along the magnetic field has to compete with pitch-angle scattering by magnetic-field fluctuations, which works to isotropize the distribution of particle pitch angles on the scattering timescale $t_{\rm scatt}$, in addition to inducing stochastic particle acceleration on the timescale $t_{\rm acc} \sim t_{\rm scatt}/\beta_a^2$, where $\beta_a$ represents the velocity (normalized by $c$) of the accelerating agents.

The fraction of energy gained in the magnetic field-aligned direction, as well as the scattering timescale, are governed by the strength of the guide field with respect to the reconnecting magnetic field. Since $t_{\rm scatt}$ increases with higher $B_g/B_0$ values, both pitch angle isotropization and stochastic particle acceleration are progressively suppressed as $B_g/B_0$ increases. 
In Figure \ref{fig7}(a), we show examples of the time evolution of the Lorentz factor (divided by $\sigma_0$) and the pitch angle cosine for particles energized in simulations with lepton magnetization $\sigma_0=16$ and different guide field strengths, namely $B_g = (1/8, 1, 8) B_0$. Particles accelerated with a low pitch angle ($\cos{\alpha} \simeq 1$) typically undergo pitch angle scattering and additional acceleration when the guide field is small ($B_g=B_0/8$) or moderate ($B_g=B_0$). In contrast, their pitch angle remains nearly unchanged after the initial acceleration when the guide field is strong ($B_g=8B_0$). 

Pitch angle isotropization following the initial acceleration stage can be evaluated computing the pitch-angle correlation function $\langle \cos{\alpha} \cos{\alpha_0} \rangle$ for particles with $|\cos{\alpha_0}| \in [0.8,1]$ and $\gamma \in [\sigma_0/2, 2 \sigma_0]$ (i.e., particles that have undergone magnetic field-aligned acceleration). Figure \ref{fig7}(b) indicates that the decorrelation time of the pitch-angle cosine is shorter as $B_g/B_0$ decreases. We can estimate the scattering timescale by fitting a decaying exponential $e^{-t/t_{\rm scatt}}$ to the pitch-angle correlation function. The resulting timescales for all the different $B_g/B_0$ simulations with $\sigma_0=16$ are presented in Figure \ref{fig7}(c). The scattering timescale roughly follows the relation $t_{\rm scatt} \sim (1+B_g^2/B_0^2) L_x/c$. Consequently, when $B_g/B_0 \lesssim 1$, pitch angle scattering operates on the advection timescale. In contrast, for $B_g/B_0 \gg 1$, a significantly larger number of light-crossing times $L_x/c$ are necessary to fully randomize the small pitch angles imprinted by the reconnection electric field.

Since $t_{\rm scatt} \gg L_x/\langle v_{\rm out} \rangle$ for $B_g/B_0 \gg 1$, in the asymptotic limit, the influence of pitch angle scattering on the pitch angle anisotropy can be neglected. Assuming that particles initially possess $\gamma_{\rm init} \sim 1$ and that all their energy gain occurs in the magnetic field-aligned direction, then $v_\parallel/c \sim (1 - 1/\gamma)^{1/2}$ for $\Delta \gamma = \gamma - \gamma_{\rm init} \gg 1$. Therefore, in this limit, the pitch angle anisotropy scales as $\langle \sin^2 \alpha \rangle \propto \gamma^{-2}$ for $1 \ll \gamma < \gamma_{\rm cut}$. Given that $\gamma_{\min \alpha} \sim \gamma_{\rm cut} \sim 4 \sigma_0$ in this regime, then $\min \langle \sin^2 \alpha \rangle \sim 1/(4 \sigma_0)^2$, which is consistent with the trend reported in Figure \ref{fig6}.

We finally note that in fully 3D magnetic reconnection, additional resonant wave interactions and instability-driven scattering centers, not permitted for $\partial/\partial z = 0$, can serve as extra isotropization channels. However, since the initial particle acceleration operates on a timescale much shorter than the scattering timescale, we anticipate the general applicability of the results presented here \citep[see][]{Comisso19}. We will explore this further in our subsequent work.

\section{Implications}\label{sec:implications} 

The concurrent particle acceleration and generation of energy-dependent pitch angle anisotropy carries several astrophysical implications. Here, we discuss some of the most important consequences.

\vspace{0.3cm}

- \emph{Synchrotron luminosity.} The synchrotron radiation power from an electron is given by $P_{\rm sync} = 2 \sigma_T c U_B \gamma^2 (v/c)^2 \sin^2 \alpha$ in the comoving frame. Here, $\sigma_T$ denotes the electron Thomson scattering cross section and $U_B$ is the magnetic energy density. Since $P_{\rm sync} \propto \sin^2 \alpha$, emission mechanisms other than synchrotron radiation can take over even in the presence of very strong magnetic fields. One such mechanism is the inverse Compton (IC) scattering of external and synchrotron photons. 
Neglecting Klein-Nishina effects on IC scattering, we can express the ratio of synchrotron to IC luminosity as
\begin{equation}
\frac{L_{\rm sync}}{L_{\rm IC}}= \left(\frac{U_{B}}{U_{\rm rad}}\right) \langle \sin^2 \alpha \rangle \, , 
\end{equation}
where $U_{\rm rad}$ is the radiation energy density, encompassing both the energy density of the external photon field and that of the synchrotron photons. Therefore, the IC luminosity is expected to be higher than the synchrotron luminosity if $\langle \sin^2 \alpha \rangle < U_{\rm rad}/U_B$. 
In cases where the pitch-angle anisotropy is most extreme, such as when $B_g/B_0 \gg 1$ and $\sigma_0 \gg 1$, most of the energy is carried by particles with $\gamma = \kappa \sigma_0$. In this scenario, the ratio of synchrotron to IC luminosity becomes 
\begin{equation}
\frac{L_{\rm sync}}{L_{\rm IC}} \sim \frac{1}{\kappa^2 \sigma_0^2} \left(\frac{U_{B}}{U_{\rm rad}}\right) \, .
\end{equation}
Under these conditions, the pitch-angle anisotropy of reconnection-accelerated particles promotes a suppression in synchrotron luminosity by a factor of approximately $\kappa^2 \sigma_0^2$ compared to the isotropic assumption. This possibility has been recently invoked to explain orphan $\gamma$-ray flares in blazars \citep{Sobacchi2021MNRAS}, the spectral energy distributions of blazars and $\gamma$-ray bursts \citep{SSB2021}, as well as the heating of magnetar magnetospheres \citep{Nattila22}.

\vspace{0.3cm}

- \emph{Spectral energy distribution.} The synchrotron spectrum generated by an anisotropic distribution of nonthermal particles can markedly differ from that of an isotropic population \citep[see ][]{Comisso20ApJL,Tavecchio2020MNRAS}.
Here, we briefly discuss the principal features of this spectrum.
We characterize the optically thin synchrotron spectrum through the energy flux, $\nu F_\nu$. 
From the scaling of the emission peak for electrons with Lorentz factor $\gamma$, occurring near the critical frequency $\nu_c = (3/2) \gamma^2 \nu_L \langle \sin \alpha \rangle \propto \gamma^{2+m/2}$, and the scaling of the power radiated by such electrons, $P_{\rm sync} = 2 \sigma_{\rm T}c U_B \gamma^2 \langle \sin^2 \alpha \rangle \propto\gamma^{2+m}$, it is straightforward to deduce the photon energy flux 
\begin{equation}
\label{eq:nuFnu}
\nu F_\nu \sim \gamma \frac{dN}{d\gamma} P_{\rm sync} \propto \nu^{(3-p+m)/(2+m/2)} \, .
\end{equation}
The usual isotropic case, $\nu F_\nu \propto\nu^{(3-p)/2}$, is recovered setting $m=0$. 
Therefore, an anisotropic pitch angle distribution can significantly harden the synchrotron spectrum for $m=m_> >0$, while it softens the spectrum for $m=m_< <0$. 
For the regime $B_g/B_0 \sim 1$, we obtained $\gamma_{\min \alpha} \sim \gamma_0$. Therefore, using characteristic values of $p_> \sim 2.2$ and $m_> \sim 2$ (see Figs. \ref{fig4} and \ref{fig6}), we obtain $\nu F_\nu \propto\nu^{0.7}$. This should be contrasted with the $\nu F_\nu \propto\nu^{0.4}$ spectrum observed when the pitch angle does not depend on particle energy. 

Another important point is that the critical synchrotron frequency of the particles with the strongest pitch angle anisotropy is reduced compared to the isotropic case. Now,
\begin{equation}
\nu_c(\gamma_{\min \alpha}) \simeq \frac{3}{2} \gamma_{\min \alpha}^2 \left(\frac{\gamma_{\min \alpha}}{\gamma_{\rm iso}} \right)^{m_>/2} \nu_L \, ,
\end{equation}
which is a factor of $(\gamma_{\min \alpha}/\gamma_{\rm iso})^{m_>/2}$ lower than the critical frequency in the isotropic scenario ($m_>=0$).
As a result, pitch-angle anisotropy in the $m_>$ range broadens the frequency range of the synchrotron spectrum in comparison to the isotropic case, while the opposite holds for the $m_<$ range. The hardening of the synchrotron spectrum due to the energy-dependent pitch-angle anisotropy spans the frequency range given by
\begin{equation}
\frac{\nu_c(\gamma_{\rm iso})}{\nu_c(\gamma_{\min \alpha})} \simeq \left(\frac{\gamma_{\rm iso}}{\gamma_{\min \alpha}} \right)^{(4+m_>)/2} \, .
\end{equation}
For the regime $B_g/B_0 \sim 1$, we estimated $\gamma_{\rm iso}/\gamma_{\min \alpha} \sim 4 \sigma_0/\gamma_0 \sim 20$. Given that typical values of $m_>$ for $B_g/B_0 \sim 1$ and $\sigma_0 \gg 1$ hover around $1.2-2.6$ (see Figure \ref{fig6}), the hardening of the synchrotron spectrum due to the pitch-angle anisotropy is anticipated to span a frequency range of $3-4$ decades. A similar frequency range was observed to undergo hardening in recent PIC simulations of magnetically dominated turbulence \citep{Comisso20ApJL}, and it was poposed as a potential explanation for the origin of the hard radio spectra observed in pulsar wind nebulae \citep{GaenslerSlane2006, Reynolds17} without the need to invoke hard ($p < 2$) particle distributions.

\vspace{0.3cm}

- \emph{Polarization.}  We evaluate the degree of linear polarization for particles following power-law distributions in both energy and pitch angle, as described by Eqs. (\ref{eq:dNdgamma}) and (\ref{eq:alphagamma}). The linear polarization degree can be calculated as 
\begin{equation} 
\Pi_{\rm lin} = \frac{\displaystyle\int G_{\rm sync}(x) \langle \sin \alpha \rangle \dfrac{dN}{d\gamma} d\gamma}{\displaystyle\int F_{\rm sync}(x) \langle \sin \alpha \rangle \dfrac{dN}{d\gamma} d\gamma} \, .
\end{equation}
Here, we have defined $F_{\rm sync}(x)=x\int_x^{\infty} K_{\frac{5}{3}}(\xi)d\xi$ and $G_{\rm sync}(x)=x K_{\frac{2}{3}}(x)$, with $x=\nu/\nu_c$ \citep{RybLig79}, where $K_{\frac{5}{3}}$ and $K_{\frac{2}{3}}$ are modified Bessel functions of orders $5/3$ and $2/3$, respectively.
Therefore, after computing the integrals we obtain
\begin{equation} 
\Pi_{\rm lin} = \frac{p+1}{p+7/3+m/3} \, .
\end{equation}
The result for an isotropic population of particles is recovered for $m = 0$. Pitch angle anisotropy reduces the degree of linear polarization in the $m = m_>$ range, while it increases it in the $m = m_<$ range. 
In the regime $B_g/B_0 \sim 1$, we obtained $\gamma_{\min \alpha} \sim \gamma_0$ and $\gamma_{\rm iso} \sim 4 \sigma_0 < \gamma_{\rm cut}$. Therefore, using $p_< \sim 3/4$ and $m_< \sim -1.3$ as typical values (see Figs. \ref{fig4} and \ref{fig6}), we obtain a linear polarization degree of $\Pi_{\rm lin} \sim 66\%$ within the range $\gamma_{\rm th} < \gamma < 0.2 \sigma$. In the range $0.2 \sigma < \gamma < 4 \sigma_0$, using $p_> \sim 2.5$ and $m_> \sim 2$ (see Figs. \ref{fig4} and \ref{fig6}), we obtain a similar linear polarization degree of $\Pi_{\rm lin} \sim 64\%$. Finally, in the range $4 \sigma_0 < \gamma < \gamma_{\rm cut}$, the degree of linear polarization increases to $\Pi_{\rm lin} \sim 75\%$. 
We obtain a much higher degree of linear polarization for $B_g/B_0 \gg 1$, where the pitch-angle anisotropy exhibits an extended $m = m_<$ range (see Figure \ref{fig5}(a)), and  $\gamma_{\min \alpha} \sim \gamma_{\rm cut} \sim 4 \sigma_0$. In this case, we have again $\Pi_{\rm lin} \sim 66\%$ within the range $\gamma_{\rm th} < \gamma < 0.2 \sigma$. However, in the range $0.2 \sigma < \gamma < 4 \sigma_0$, using $p_< \sim 5$ and $m_< \sim -2$ as typical values (see Figs. \ref{fig4} and \ref{fig6}), we obtain a significantly higher linear polarization degree of $\Pi_{\rm lin} \sim 90\%$.

Radiation from an anisotropic electron distribution could have a significant circular polarization degree. The circular polarization degree for such a distribution within a plasma consisting mainly of electrons and ions can be evaluated as \citep{Sazonov72} 
\begin{eqnarray}
\Pi_{\rm cir} &=& \frac{1}{\gamma_0} \frac{p+1}{p+7/3} \frac{\Gamma\left(\dfrac{3p+4}{12}\right) \Gamma\left(\dfrac{3p+8}{12}\right)}{\Gamma\left(\dfrac{3p-1}{12}\right) \Gamma\left(\dfrac{3p+7}{12}\right)} \nonumber\\& &\times  \frac{(2+p)\cot \theta - \sin \theta g(\gamma_0, \theta)}{p} \, ,
\end{eqnarray}
where $\Gamma$ is the Gamma function, and $\theta$ is the angle between the direction of the magnetic field and the line of sight. The viewing angle is equivalent to the pitch angle of the electrons that primarily contribute to the observed radiation. Therefore, by setting $\cos \alpha = \cos \theta$, the function $g$ can be expressed as
\begin{equation} 
g(\gamma_0, \theta) = \left( \frac{1}{f(\gamma, \cos \alpha)}  { {\frac{\partial f(\gamma,\cos \alpha)}{\partial \cos \alpha}}} \right)_{\gamma=\gamma_0} \, .
\end{equation}  
The effect of the pitch-angle anisotropy is included in $g(\gamma_0,\theta)$, which is zero for isotropic distributions. Therefore, a substantial level of circular polarization, with $\Pi_{\rm cir} \gg 1/\gamma_0$, can be attained with an anisotropic pitch angle distribution.

\vspace{0.3cm}

- \emph{Fast cooling.}  Electrons with Lorentz factor $\gamma \gg 1$ cool via synchrotron radiation on the timescale $t_{\rm{cool}} = {m_e c}/({2 \sigma_T U_B \gamma \sin^2 \alpha})$. Particles are strongly cooled if they radiate a significant fraction of their energy in a timescale shorter than $L_x/\langle v_{\rm out} \rangle$. The Lorentz factor of a particle that cool in such timescale can be expressed as 
\begin{equation} \label{gamma_cool}
\gamma_{\rm{cool}} = \frac{1}{2 \ell_B} \frac{1}{\sin^2 \alpha} \bigg(1+\frac{B_g^2}{B_0^2}\bigg)^{-1/2} \, ,
\end{equation}
where 
\begin{equation} \label{ell_B}
\ell_B = \frac{\sigma_T U_B L_x}{m_e c^2}
\end{equation}
is the magnetic compactness parameter. Particles with $\gamma > \gamma_{\rm{cool}}$ are in a fast cooling regime, while particles with $\gamma < \gamma_{\rm{cool}}$ are in a slow cooling regime. The particle spectrum is affected by cooling on the timescale $L_x/\langle v_{\rm out} \rangle$ if it extends above $\gamma_{\rm{cool}}$. 
For $B_g/B_0 \gtrsim  1$, particle injection up to $\gamma = \gamma_0 \simeq \kappa \sigma_0$ is not affected by cooling unless 
\begin{equation} 
\ell_B \gtrsim \frac{B_0}{2 \kappa \langle \sin^2 \alpha \rangle \sigma_0 B_g} \sim 0.1 \sigma_0  \frac{B_0}{B_g} \, .
\end{equation}
Under conditions that favor the strongest pitch-angle anisotropy, namely $B_g/B_0 \gg 1$ and $\sigma_0 \gg 1$, the electron energy spectrum extends up to $\gamma_{\rm cut} \sim 4 \sigma_0$, and the particle spectrum is affected by synchrotron cooling only if $\ell_B \gtrsim 2 \sigma_0  {B_0}/{B_g}$. 

Marginally fast cooling \citep{Daigne11,Beniamini18,Xu2018ApJ} and pitch-angle anisotropy \citep{Lloyd2000,YangZhang18,Goto22} have been proposed as key factors shaping the prompt emission spectrum of $\gamma$-ray bursts. When considering an energy-dependent pitch-angle anisotropy, the resulting cooled particle distribution is nontrivial as the pitch angle is regulated by the interplay between particle acceleration, scattering, and cooling. In general, due to the $t_{\rm cool} \propto 1/\sin^2 \alpha$ dependence, synchrotron cooling is biased toward cooling particles with larger pitch angles. With the energy-dependence of pitch-angle anisotropy given in Eq. (\ref{eq:alphagamma}), $t_{\rm cool} \propto \gamma^{-(1+m)}$. Therefore, within the $m_<$ range, cooling becomes stronger for lower-energy particles when $m_< <-1$. This leads to a hardening of the particle energy spectrum in the range controlled by the $m_<$ slope \citep{Comisso21}. The faster cooling of lower-energy particles ceases when $m_< \geq -1$. This suggests that strong synchrotron cooling tends to induce and maintain a pitch angle anisotropy with $m_< \sim -1$.

\vspace{0.3cm}

- \emph{Synchrotron burnoff limit.}  In the realm of ideal plasmas, where the electric field is smaller, or at most equal, to the magnetic field, it is widely acknowledged that astrophysical sources cannot emit synchrotron radiation above $h \nu_{\rm burnoff} \simeq 160 \, {\rm{MeV}}$ in their rest frame \citep{Guilbert83,deJager96}. However, this constraint does not apply when dealing with particle distributions characterized by small pitch angles. On general grounds, the Lorentz factor for which the radiation reaction force $F_{RR}^{\rm{sync}}  =  2 \sigma_T \gamma^2  U_B \sin^2 \alpha$ for $\gamma \gg 1$ balances the accelerating force $F_{\rm{acc}} = e E$, known as the radiation reaction limit, is given by
\begin{equation} \label{gamma_rad}
\gamma_{\rm{rad}} = \left(\frac{e E}{2 \sigma_T U_B \sin^2 \alpha}\right)^{1/2} \, ,
\end{equation}
in the comoving frame. 
The critical synchrotron photon energy emitted by a population of particles limited by radiation reaction is then given by
\begin{equation} \label{}
h \nu_{\rm rad} = \frac{3}{2} \hbar \gamma_{\rm{rad}}^2 \omega_L {\langle \sin \alpha \rangle} \simeq \frac{16}{\langle \sin \alpha \rangle} \frac{B_0^2}{B_0^2 + B_g^2} {\rm{MeV}} \, ,
\end{equation}
where $\hbar$ is the reduced Planck constant, and we used $E_{\rm rec} = \langle R_{\rm rec} \rangle B_0$ as the relevant electric field. The maximum synchrotron photon energy depends only on the ratio $B_g/B_0$ and the pitch-angle anisotropy, which is modulated by the lepton magnetization $\sigma_0$. Given that $\langle \sin \alpha \rangle$ can reach very low values, there is no strict upper limit of $160 \, {\rm{MeV}}$. For $B_g/B_0 \gtrsim  1$, taking particles at $\gamma = \gamma_0$, one has
\begin{equation} \label{}
h \nu_{\rm rad} \simeq  3 \sigma_0 {\left( {\frac{B_0}{B_g}} \right)^2} {\rm{MeV}}  \, ,
\end{equation}
Therefore, for $\sigma_0 > 50 (B_g/B_0)^2$, reconnection emits synchrotron radiation above the ideal limit $h \nu_{\rm burnoff} \simeq 160 \, {\rm{MeV}}$. This is essential for explaining the emissions observed from some of the most extreme astrophysical accelerators, such as the $\gamma$-ray flares from the Crab Nebula \citep{BuhlerBlandford14}.

\vspace{0.3cm}

- \emph{Beaming.} Energy-dependent pitch-angle anisotropy can result in highly beamed radiation aligned with the magnetic field direction. 
This beaming effect is fundamentally distinct and anticipated to be more prevalent than the Doppler beaming associated with the bulk motions in the reconnection layer \citep{Gia09,Nalewajko11,Petropoulou16}. For a plasma with a general composition, and by applying Eq. (\ref{eq:vout}) for an arbitrary plasma magnetization, the maximum Lorentz factor of the bulk flow within the reconnection layer is given by 
\begin{equation}
\Gamma_{\rm{bulk, max}} \simeq   \left(\frac{1+\sigma_g+\sigma_0}{1+\sigma_g}\right)^{1/2} \, .
\end{equation}
 This implies that even a moderate guide field strength $B_g \gtrsim B_0/3$ or an overall magnetization $B_0^2/[4\pi c^2(m_e n_0 + m_i n_i)] = \sigma_0/(1+\sigma_0/\sigma_{0,i}) \lesssim 1$ results in  $\Gamma_{\rm{bulk, max}} \sim 1$, effectively eliminating the possibility of bulk outflow beaming.
On the other hand, the beaming associated with the pitch-angle anisotropy only requires high lepton magnetization, $\sigma_0 \gg 1$. In this case, the alignment of particle momentum with the local magnetic field reaches a maximum at a Lorentz factor associated with the lepton magnetization ($\gamma_{\min \alpha} \ll \sigma_0$ for $B_g/B_0 \ll 1$, $\gamma_{\min \alpha} \sim \kappa \sigma_0$ for $B_g/B_0 \sim 1$, and $\gamma_{\min \alpha} \sim 4 \sigma_0$ for $B_g/B_0 \gg 1$). 
As particles reach higher Lorentz factors, well above $\gamma_{\min \alpha}$, their alignment with the magnetic field gradually diminishes due to pitch-angle scattering off magnetic field fluctuations, which occurs on a timescale $t_{\rm scatt}$ much longer than the timescale 
\begin{equation}
t_{\min \alpha} = \frac{\gamma_{\min \alpha}}{\langle R_{\rm rec}\rangle}  \frac{m_e c}{e B_0} \, 
\end{equation}
associated with the initial particle acceleration that leads to small pitch angles. Therefore, particles with $\gamma \sim \gamma_{\min \alpha}$ are expected to predominantly emit radiation in the direction of $B_0$ when $B_g/B_0 \ll 1$ \citep{Cerutti2012ApJL} and in the direction of $B_g$ when $B_g/B_0 \gtrsim 1$ \citep{Comisso20ApJL}. As the particles' Lorentz factor increases to values $\gamma \gg \gamma_{\min \alpha}$, their emission becomes increasingly isotropic, and particularly pronounced in the direction perpendicular to the guide field $B_g$ when $\cos \alpha \simeq 0$, since this stage of acceleration is governed by the ideal-MHD electric field (a similar picture holds for relativistic turbulence \citep{Comisso19}).

When radiation emission occurs on a timescale longer than that required for imprinting the pitch-angle anisotropy but shorter than the scattering timescale, i.e., $t_{\min \alpha} \ll t_{\rm cool} \ll t_{\rm scatt}$, the emitted radiation, either via synchrotron or inverse Compton scattering, is beamed into a solid angle $\Delta \Omega \sim \langle \sin^2 \alpha \rangle$ around the local magnetic field. The extent of $\Delta \Omega$ is tied to the particle's energy due to the energy-dependent nature of pitch-angle anisotropy, as outlined in Eq. (\ref{eq:alphagamma}). For $B_g/B_0 \gtrsim  1$, one has
\begin{equation} 
\Delta \Omega \sim \frac{1}{(\kappa \sigma_0)^2} \, ,
\end{equation}
for particles at $\gamma = \gamma_0$. In the regime $B_g/B_0 \gg 1$, the beaming effect intensifies as the Lorentz factor increases, culminating in a maximum beaming with a solid angle of $\Delta \Omega \sim {1}/{(4 \sigma_0)^2}$ at the cutoff energy. 

In the presence of strong beaming, the emission from a reconnecting current sheet can only be observed if the local magnetic field is closely aligned with the observer's line of sight. This effect can lead to rapid variability of the emission. 
When multiple active radiation beams are present and the beams are distributed isotropically, the variability timescale $\delta T$ for an emission event spanning a timescale $T$ depends on the pitch angle as
\begin{equation} 
\frac{\delta T}{T} \sim N_{\alpha} \langle \sin^2 \alpha \rangle \, ,
\end{equation}
where $N_{\alpha}$ represents the number of active beams. This ``lighthouse effect'' has been exploited in \citet{Sob23} to potentially explain the ultrafast variability of $\gamma$-ray flares from active galactic nuclei. In this scenario, the variability resulting from pitch-angle anisotropy is frequency-dependent, in contrast to the achromatic variability associated with ultrarelativistic bulk motions \citep{Lyutikov06,NarayanKumar09,Lazar09,Gia09}.

\vspace{0.3cm}

- \emph{Temperature anisotropy.} Pitch-angle anisotropy, where electrons' velocities are more aligned with magnetic field lines, results in an anisotropic electron temperature, with $T_{e,\parallel} > T_{e,\perp}$.  The temperatures $T_{e,\parallel} = P_{e,\parallel}/k_B n_e$ and $T_{e,\perp} = P_{e,\perp}/k_B n_e$ are associated with electron pressures parallel and perpendicular to the magnetic field direction, respectively, with $P_{e,\parallel}$ and $P_{e,\perp}$ representing the corresponding pressures. 
In cases where the pitch angle anisotropy is most extreme, such as when $B_g/B_0 \gg 1$ and $\sigma_0 \gg 1$, the electron temperature anisotropy intensifies, resulting in $T_{e,\parallel} \gg T_{e,\perp}$. 
The sustainability of high levels of temperature anisotropy depends on the extent to which it's allowed by anisotropy-driven instabilities. 
For $T_{e,\parallel} \gg T_{e,\perp}$, the anisotropy might be limited by the onset of the firehose instability \citep{Gary93}, which is triggered when enough pressure anisotropy is generated to counteract the magnetic tension.
The threshold for triggering the parallel firehose instability in an electron-ion plasma is given by \citep{Barnes73}
\begin{equation} 
\frac{P_{e,\perp}}{P_{e,\parallel}} < 1- \frac{2}{\beta_{e,\parallel}}  \, ,
\end{equation}
where $\beta_{e,\parallel} = 2 \theta_{e,\parallel}/\sigma_e$ and $\theta_{e,\parallel} = k_B T_{e,\parallel}/m_e c^2$. In pair plasmas, where both electrons and positrons contribute equally to the instability, the threshold condition is modified by $\beta_{e,\parallel} \rightarrow 2 \beta_{e,\parallel}$.
In the magnetically dominant regime under our current focus, $\beta_{e,\parallel} \ll 1$, which enables the maintenance of strong temperature anisotropies. 

Proper modeling of temperature anisotropy is critical for correctly interpreting synchrotron radiation from astrophysical plasmas, and it can significantly impact conclusions drawn from observations when compared to models that assume isotropic particle distributions. Recent work by \citet{Galishnikova23} has indeed shown that temperature anisotropy in accreting plasmas significantly affects the interpretation of mm-wavelength observations, such as those related to black hole imaging for Sgr A* and M87* \citep{EHT_M87_19,EHT_SgrA_22}.

\section{Conclusions}\label{sec:conclusions} 

In this article, we have investigated the concurrent generation of energetic particles and pitch-angle anisotropy by magnetic reconnection in magnetically dominated plasmas. 
Using rigorous first-principles PIC simulations, we demonstrated that reconnection-driven particle acceleration imprints an \emph{energy-dependent pitch-angle anisotropy} and gives rise to broken power laws in both the particle energy spectrum and the pitch angle anisotropy. The properties of reconnection-accelerated particle distributions depend strongly on the relative strength of the guide field compared to the reconnecting magnetic field ($B_g/B_0$), in addition to the lepton magnetization ($\sigma_0$). These are pivotal in shaping the particle energy spectrum and pitch angle anisotropy.

The break Lorentz factor $\gamma_0$, which separates the two power-law ranges in the particle energy spectra, depends on lepton magnetization and exhibits minimal sensitivity to the guide field's strength. This break Lorentz factor is a significant fraction of the lepton magnetization, specifically $\gamma_0 = \kappa \sigma_0$, with $\kappa \simeq 0.2$. For $\sigma_0 \gg 1$, this results in extended pre-break ranges.
Below $\gamma_0$, the particle energy spectrum is governed by particle injection, yielding a remarkable hard slope ($p_< \simeq 1/2 - 3/4$) that is robust to variations in guide field strength and magnetization. In contrast, the power-law range above $\gamma_0$ exhibits significant sensitivity to the guide field's strength ($p_> \sim 1.5 - 5$ for $\sigma_0 \gg 1$), in addition to its dependence on magnetization. The extension of this second power-law range also hinges on the guide field strength, with $B_g/B_0 \gg 1$ leading to a high-energy cutoff Lorentz factor $\gamma_{\rm cut}$ that is proportional to $\sigma_0$, while $B_g/B_0 \ll 1$ results in $\gamma_{\rm cut}$ approaching the Hillas limit of the reconnection layer.

The energy-dependent pitch-angle anisotropy is characterized by the Lorentz factor at which the pitch angle $\alpha$ reaches its minimum, $\gamma_{\min \alpha}$, and the Lorentz factor at which particles return to a state close to pitch-angle isotropy after being energized, $\gamma_{\rm iso}$. The existence of the latter depends on the strength of the guide field. Essentially, we can distinguish three regimes, depending on the relative strength of the guide field in comparison to the reconnecting magnetic field.
For $B_g/B_0 \ll 1$ (moderate anisotropy), $\gamma_{\min \alpha} \ll \sigma_0$, $\gamma_{\rm iso} \sim \sigma_0 \ll \gamma_{\rm cut}$, and the deviation of $\langle \sin^2 \alpha \rangle$ from isotropy is modest though noticeable. 
For $B_g/B_0 \sim 1$ (strong anisotropy), $\gamma_{\min \alpha} \sim \gamma_0$, $\gamma_{\rm iso} \sim 4 \sigma_0 < \gamma_{\rm cut}$, the negative slope ($m_<$) and positive slope ($m_>$) of the energy dependence exhibit steep power laws ($|m| \gtrsim 1$), and the pitch-angle anisotropy can reach $\min \langle \sin^2 \alpha \rangle \sim 1/(\kappa \sigma_0)^2$. 
For $B_g/B_0 \gg 1$ (extreme anisotropy), $\gamma_{\min \alpha} \sim \gamma_{\rm cut} \sim 4 \sigma_0$, the negative slope ($m_<$) dominates the energy range, and the pitch-angle anisotropy can reach $\min \langle \sin^2 \alpha \rangle \sim 1/(4 \sigma_0)^2$.

We discussed several astrophysical implications of the concurrent particle acceleration and generation of energy-dependent pitch-angle anisotropy. 
In particular: \\
(1) Pitch-angle anisotropy progressively suppresses synchrotron luminosity with increasing guide field, reaching roughly a factor of $\kappa^2 \sigma_0^2$ when $B_g/B_0 \gg 1$ and $\sigma_0 \gg 1$. \\
(2) The synchrotron energy flux for an energy-dependent pitch angle distribution is given by $\nu F_\nu \propto \nu^{(3-p+m)/(2+m/2)}$, and thus it hardens in the $m=m_>$ range compared to an energy-independent pitch angle, while softening occurs in the $m=m_<$ range. For $B_g/B_0 \sim 1$, we estimate the hardening of the synchrotron spectrum spans approximately $3-4$ decades in frequency. \\ 
(3) The degree of linear polarization is given by $\Pi_{\rm lin} = (p+1)/(p+7/3+m/3)$, and thus pitch angle anisotropy reduces it in the $m=m_>$ range, while increasing it in the $m=m_<$ range. The anisotropic pitch angle distribution can also increase the degree of circular polarization to $\Pi_{\rm cir} \gg 1/\gamma_0$. \\ 
(4) Pitch-angle anisotropy enforces stricter requirements for entering the fast cooling regime. When $B_g/B_0 \gtrsim 1$, particle injection remains unaffected by cooling unless the magnetic compactness parameter $\ell_B \gtrsim 0.1 \sigma_0 B_0/B_g$. For highly anisotropic distributions, strong synchrotron cooling is expected to induce a pitch angle anisotropy with $m_< \sim -1$. \\
(5) Particles accelerated by reconnection can emit radiation at energies far exceeding the burnout limit, $h \nu_{\rm burnoff} \simeq 160 \, {\rm{MeV}}$. When $B_g/B_0 \gtrsim 1$, a substantial fraction of radiation can be emitted up to $h \nu_{\rm rad} \simeq 3 \sigma_0 (B_0/B_g)^2 {\rm{MeV}}$. \\
(6) Energy-dependent pitch-angle anisotropy can result in highly beamed radiation aligned with the magnetic field direction, leading to fast variability. In this scenario, the variability resulting from pitch-angle anisotropy is frequency-dependent, in contrast to the achromatic variability associated with the Doppler beaming from bulk motions. \\ 
(7) Pitch-angle anisotropy results in an anisotropic electron temperature, with $T_{e,\parallel} > T_{e,\perp}$. In the magnetically dominated regime, very high levels of temperature anisotropy can be sustained against plasma instabilities.

In magnetically dominated collisionless plasmas, pitch angle anisotropy is anticipated as the norm rather than the exception. 
Knowledge of both the particle energy spectrum and pitch-angle anisotropy is necessary to understand the radiation signatures emitted by energized particles. 
This work has provided a first-principles description of nonthermal particle acceleration in reconnection, encompassing both the particle energy spectrum and energy-dependent pitch-angle anisotropy.
This will enables us to develop reliable radiative models with improved predictive power compared to current approaches that neglect pitch angle anisotropy.

\section*{Acknowledgments}
We acknowledge fruitful discussions with Emanuele Sobacchi and Lorenzo Sironi. This research is supported by the NASA ATP award 80NSSC22K0667. We acknowledge computing resources from Columbia University's Shared Research Computing Facility project, which is supported by NIH Research Facility Improvement Grant 1G20RR030893-01, and associated funds from the New York State Empire State Development, Division of Science Technology and Innovation (NYSTAR) contract C090171.

\bibliography{rec_pitchangle}{}
\bibliographystyle{aasjournal}

\end{document}